\documentclass[preprint,showkeys,amsfonts,amsmath,amssymb]{revtex4}
\usepackage{graphicx}
\usepackage{dcolumn}
\usepackage{subfigure}
\usepackage{epstopdf}
\usepackage{boondox-cal}
\usepackage{enumitem}
\usepackage{bm}

\begin{document}

\title{Scattering and absorption of a charged massive scalar field by a Reissner-Nordstr\"om black hole surrounded by perfect fluid dark matter}

\author{Hai Huang}
\thanks{Corresponding author: haihuang@gyu.edu.cn}
\affiliation{College of Mechanical Engineering, Guiyang University, Guiyang, Guizhou 550005, China}
\author{Xudong Sun}
\affiliation{Department of Theoretical Physics, Kunming University of Science and Technology, Kunming, Yunnan 650091, China}
\author{Juhua Chen}
\thanks{Corresponding author: jhchen@hunnu.edu.cn}
\affiliation{Department of Physics, Hunan Normal University, Changsha, Hunan 410081, China}

\begin{abstract}
We study the scattering of charged massive particles impinging on a Reissner-Nordstr\"{o}m (RN) black hole immersed in perfect fluid dark matter (PFDM). We obtain an approximation of absorption cross section in the low-frequency regime via the matching method. In the high-frequency regime, we derive the weak-field deflection angle up to second order. The numerical results are in excellent agreement with classical approximation and glory scattering. The effects of dark matter, particle charge, and mass upon scattering and absorption are examined in detail. The results show that as the dark matter parameter $\lambda$ increases, the absorption cross section of the black hole is strongly suppressed, and its high-frequency limit depends only on the black hole charge $Q$ and $\lambda$. The scattering cross section also decreases overall. In the superradiant regime, the amplification factor of the PFDM black hole is much larger than that of the RN black hole. Finally, we discuss the behavior of the absorption cross section as $\omega/m\rightarrow1$, as well as the scattering cross section at small scattering angles.
\end{abstract}


\maketitle

\section{Introduction}

Black hole (BH) scattering is one of the central problems in studying wave propagation and gravitational interactions in curved spacetime. In early seminal work, S\'{a}nchez systematically investigated the elastic scattering of scalar waves by BHs~\cite{Sanchez1978}, followed by Matzner \textit{et al.}, who introduced the well-known glory scattering mechanism to describe the backward enhancement of scattered waves~\cite{Matzner1985}. These studies established the fundamental framework for analyzing BH scattering using partial wave methods, phase shift analysis, and geometric optics approximations. Since then, extensive theoretical efforts have been devoted to BH scattering in different spacetimes and for different types of fields~\cite{Crispino2009,Dolan2006,Leite2018,Dolan2013,Benone2014,Chen2013,Benone2016,Aoude2023,Firrotta2025}. In particular, scattering theory has been further developed to incorporate absorption effects into a unified description, showing that both the scattering and absorption cross sections are highly sensitive to the background geometry and to the structure of the effective potential. More recently, BH scattering has also been reformulated within modern scattering amplitude and effective field theory approaches. In this framework, compact objects are described by worldline degrees of freedom, and classical observables, such as the impulse, scattering angle, and absorptive corrections, can be extracted from relativistic amplitudes. Horizon absorption is encoded through dissipative worldline operators, whose Wilson coefficients or spectral densities are matched to the corresponding BH absorption cross sections. This provides a complementary interpretation of the absorption probability obtained from the partial wave method~\cite{Jones2024,Goldberger2006,Goldberger2006a,Porto2008,Kosower2019}.

On the other hand, increasing evidence suggests that astrophysical BHs are typically embedded in dark matter halos, deviating from ideal vacuum solutions. To describe such effects, an important class of solutions originates from the work of Kiselev~\cite{Kiselev2003,LiYang2012}, who constructed spherically symmetric BH solutions surrounded by matter fields within general relativity. Strictly speaking, these solutions do not correspond to an isotropic perfect fluid, as pointed out by Visser~\cite{Visser2020}. Nevertheless, they are still commonly referred to as PFDM backgrounds in the literature.
Motivated by these developments, extensive studies have been carried out on BHs surrounded by PFDM. For instance, Hou~\cite{Hou2018} investigated the shadow of rotating BHs in PFDM backgrounds; Haroon ~\cite{Haroon2019} analyzed the deflection angle and shadow in the presence of a cosmological constant; Rizwan~\cite{Rizwan2019} explored precession frequencies to distinguish PFDM BHs from naked singularities. Moreover, PFDM also affects perturbative and quantum properties of BHs. B\'ecar ~\cite{Becar2024} studied massive scalar perturbations in such backgrounds, while Papnoi ~\cite{Papnoi2024} examined thermodynamics and lensing effects. Sadeghi~\cite{Sadeghi2024} discussed implications for the weak gravity conjecture and cosmic censorship, and Ma~\cite{Ma2024} extended PFDM to nonlinear electrodynamics scenarios. In addition, Pedraza~\cite{Pedraza2026} investigated classical and semiclassical scattering by regular BHs immersed in PFDM and found that dark matter significantly modifies diffraction patterns at small scattering angles. Tovar~\cite{Tovar2025} analyzed the dependence of quasinormal modes on both BH parameters and PFDM parameters. Overall, PFDM parameters systematically modify the photon sphere structure, effective potential barrier, and wave propagation properties. Recent studies have provided preliminary observational support for the phenomenological viability of the PFDM model. By fitting the dark matter contribution to the rotation curves of representative galaxies from the SPARC database, it was shown that the Power-Law and Logarithmic solutions can describe different radial regions of galactic halos. The Power-Law solution is more suitable for the inner and intermediate regions of the rotation curve, whereas the Logarithmic solution captures the large-radius asymptotic behavior~\cite{Kuncewicz2025}. 

However, compared to studies of shadows, lensing, and quasinormal modes, scattering in PFDM backgrounds remains relatively unexplored. In particular, for charged and massive particles or scalar fields, the behavior of scattering cross sections, reflection amplitudes, and glory structures in PFDM spacetimes has not yet been systematically investigated. By contrast, similar problems have been studied in other modified gravity or non-vacuum BH backgrounds. For example, Li \textit{et al.} analyzed the absorption and scattering of charged scalar waves by charged Horndeski BHs, revealing nontrivial couplings between BH parameters and field properties~\cite{Li2025}. This suggests that extending scattering theory to PFDM backgrounds is both natural and necessary.
In this work, we investigate the scattering of charged massive particles by a charged BH immersed in PFDM. We focus on how the PFDM parameter $\lambda$, electromagnetic interactions, and particle mass affect scattering properties. Our results provide new insights into wave propagation in PFDM backgrounds and offer potential observational signatures for distinguishing different dark matter models.

This paper is arranged as follows. In Sec. II we provide the background of PFDM and derive the expressions for absorption and scattering cross sections. In Secs. III and IV we obtain the low- and high-frequency limits of the cross section, respectively. In Sec. V we present a selection of numerical results. We conclude with our final remarks and discussion in Sec. VI. We assume $c=G=\hbar=1$.

\section{The scalar field} 
Consider a charged BH is immersed in a background of PFDM \cite{Das,Das2,Zhaoyi,Anish,Xu2018}. The action can be written in the following form
\begin{equation}
S=\int \textrm{d}^4 x \sqrt{-g}\left(\frac{R}{16\pi G}-\frac{1}{4} F^{\mu v}F_{\mu v}+L_{DM}\right)  \label{action}.
\end{equation}
where $R$ is the Ricci scalar and $G$ is the Newton's gravitational constant. $F_{\mu \nu}$ is the electromagnetic field strength tensor related to 4-vector potential $A_\mu$, $F_{\mu \nu} = \partial_{\mu} A_{\nu} - \partial_{\nu} A_{\mu} $ and $L_{DM}$ gives the Lagrangian density for the perfect fluid dark matter. Extremizing the action gives the Einstein field equations
\begin{equation}
R_{\mu \nu}- \frac{1}{2}g_{\mu \nu}R=8 \pi G(T^{M}_{\mu \nu}-T^{DM}_{\mu \nu}) \label{equation}.
\end{equation}
The energy-momentum tensor for the electromagnetic field and the dark matter take the form
\begin{equation}
(T^{\mu}_{\nu})^M=\frac{Q^2}{8\pi G r^4}\textrm{diag}(-1,1,1,1);\hspace{2em}(T^{\mu}_{\nu})^{DM}=\textrm{diag}(-\rho,P_r,P_\theta,P_\varphi)\label{energy-momentum tensor}.
\end{equation}
with $P_r =-\rho$, $P_\theta = P_\varphi = P$. Where $Q$ is the BH charge, $\rho$ is the energy density and $P_r$, $P_{\theta}$ and $P_{\varphi}$ are pressures of PFDM in the three directions respectively.

This theory corresponding to the following metric of a static, spherically symmetric, charged BH in PFDM which is asymptotically Minkowski
\begin{equation}
\textrm{d}{s^2} = -f(r)\textrm{d}{t^2}+{f^{-1}}(r)\textrm{d}{r^2}+{r^2}\textrm{d}{\Omega^2}\label{element}.
\end{equation}
The lapse function $f(r)$ takes the form
\begin{equation}
f(r) =1-\frac{2M}{r}+\frac{Q^2}{r^2}+\frac{\lambda}{r} \textrm{ln} \frac{r}{|\lambda|}\label{element2}.
\end{equation}
In the lapse function, $M$ is the BH mass, $\lambda$ is the dark matter parameter, which related to the PFDM energy density $\rho$ as $\rho=\frac{\lambda}{8 \pi r^3}$ \cite{Das}, $\lambda$ can be positive or negative, here we only disscuss positive values and  $\lambda$ allowed maximum value for $\lambda=2M$\cite{Xu2018}. In the limit  $\lambda\rightarrow0$, Eq.~(\ref{element2}) coincides with the RN BH solution. In the case where $Q=0$, the metric function describes the Schwarzschild BH immersed in PFDM .

Consider a charged massive scalar field with mass $m$ and charge $q$, propagating in a static RN spacetime surrounded by PFDM. The charged massive scalar field $\Phi$ are
minimally connected to gravity as
\begin{equation}
(\bigtriangledown_{\mu}-iq A_{\mu})(\bigtriangledown^{\mu}-iq A^{\mu})\Phi-m^2\Phi=0\label{kg},
\end{equation}
it can be written as
\begin{equation}
\frac{1}{\sqrt{-g}}(\partial_{\mu}-iq A_{\mu})[\sqrt{-g}g^{\mu \nu}(\partial_{\nu}-iq A_{\nu})\Phi]-m^2\Phi=0\label{kg2}.
\end{equation}
where $\bigtriangledown_{\mu}$ denotes the covariant derivative, $g^{\mu \nu}$ is the contravariant metric components, $g$ is the metric determinant, $A_{\mu}$ is the electromagnetic four-potential. In static spherically symmetric spacetime, the electromagnetic four-potential has only the time component non-zero. By solving the Maxwell equations $\bigtriangledown_{\mu}F^{\mu \nu}=0$ in curved spacetime. We have
\begin{equation}
A_{\mu}=(-Q/r,0,0,0)\label{A}.
\end{equation}
 Here we shall be interested in monochromatic wave-like solutions. This equation can be solved using a separation of variables ansatz \cite{Brill1972}, namely,

\begin{equation}
\Phi_{\omega l}=\frac{\psi_{\omega l}(r)}{r} \textrm{Y}_{lm}(\theta,\varphi)e^{-i \omega t}\label{Phi omega l}.
\end{equation}
Substituting Eq.~(\ref{Phi omega l}) into Eq.~(\ref{kg}), we obtain the following radial equation:
\begin{equation}
\left(\frac{\textrm{d}}{\textrm{d}r_*^2}-V_{\textrm{eff}}(r)\right)\psi_{\omega l}(r)=0\label{radial},
\end{equation}
The above equation is Schr\"{o}dinger-like, in which $r_*$ is the tortoise coordinate, defined through $\textrm{d}r_*=f(r)^{-1}\textrm{d}r$, with the effective scattering potential given by
\begin{equation}
V_{\textrm{eff}}(r)=-(\omega+qA_0(r))^2+f(r)\left(m^2+\frac{l(l+1)}{r^2}+\frac{1}{r}\frac{\textrm{d}f(r)}{\textrm{d}r}\right)\label{potential2}.
\end{equation}
To solve Eq.~(\ref{radial}), it is essential to impose appropriate boundary conditions. In this context, here we will be interested in the purely incoming waves from the past null infinity. The effective potential has a local maximum in the intermediate region of $r$ and tends to $-\kappa_H^2=V_{\textrm{eff}}(r)(r\rightarrow r_h)$ and $-k^2=V_{\textrm{eff}}(r\rightarrow +\infty)$ at the event horizon and spatial infinity, respectively. The asymptotic forms of the solution can be written as

\begin{equation}
\label{boundary}
\psi_{\omega l}(r) \approx
\begin{cases}
T_{\omega l}e^{ - i\kappa_H r_*} ,& \textrm{for} \:\:\:\: r_{\ast}\rightarrow -\infty \:\:\:\:(r\rightarrow r_h), \\
e^{ - ik r_*}+R_{\omega l}e^{ik r_*}  ,& \textrm{for} \:\:\:\: r_{\ast}\rightarrow +\infty\:\:\:\:(r\rightarrow +\infty), 
\end{cases}
\end{equation}
where $\kappa_H=\omega-qQ/r_h$ and $k=\sqrt{\omega^2-m^2}$, we usually consider scattering states $\omega>m$. These boundary conditions correspond to an incident wave of amplitude 1 from spatial infinity giving rise to a reflected wave of amplitude $R_{\omega l}$ and a transmitted wave of
amplitude $T_{\omega l}$ at the horizon. Using the relationship of Wronskian, $W=-2i\kappa_H |T_{\omega l}|^2$ at the horizon and  $W=2ik( |R_{\omega l}|^2-1)$ at infinity. It is straightforward to show that
 \begin{equation}
|R_{\omega l}|^2+\frac{\kappa_H }{k}|T_{\omega l}|^2=1\label{relation}.
\end{equation}
The phase shift $\delta_l$ is defined by
\begin{equation}
e^{2i\delta_l}=(-1)^{l+1}R_{\omega l}\label{phase}.
\end{equation}
Based on the quantum mechanics theory, the total absorption cross section can be written as:
\begin{equation}
\sigma _{abs}=\frac{\pi}{\omega^2 v^2}\sum\limits_{l=0}^\infty(2l + 1)(1-|R_{\omega l}|^2)\label{abs2},
\end{equation}
where $v=\sqrt{1-m^2/\omega^2}$ is a dimensionless parameter and $\omega v=k$. The differential scattering cross section for static and spherically spacetimes is given by
\begin{equation}
\frac{\textrm{d}\sigma}{\textrm{d}\Omega}=|f(\theta)|^2\label{scattering cross section},
\end{equation}
where the scattering amplitude $f(\theta)$ reads\cite{Crispino2009}
\begin{equation}
f(\theta)=\frac{1}{2i \omega v}\sum\limits_{l=0}^\infty (2l+1)[e^{2i \delta_l}-1]P_l(cos \theta)\label{scattering amplitude}.
\end{equation}

\section{Low frequency regime}

We focus on the low-frequency regime and assume that the logarithmic long-range contribution generated by $\lambda$ can be treated perturbatively in the far region. In this approximation, the leading far-zone solution may be matched to the standard $l=0$ Coulomb wave.

\subsection{Near horizon solution}

By introducing the transformation $\psi_{\omega l}=r\phi$, the radial equation (\ref{radial}) can be rewritten as
\begin{equation}
\frac{f(r)}{r^2}\frac{\textrm{d}}{\textrm{d}r}\left(r^2 f(r)\frac{\textrm{d}\phi}{\textrm{d}r}\right)+\Big[\big(\omega-\frac{qQ}{r}\big)^2 - f(r)\Big(m^2 +\frac{l(l+1)}{r^2}\Big)\Big]\phi=0.\label{Near-horizon solution}
\end{equation}
For region I $(r\approx r_h)$, the equation at the horizon simplifies to a constant coefficient second-order equation
\begin{equation}
\frac{\textrm{d}^2\phi}{\textrm{d}r_*^2}+\kappa_H^2\phi_{\rm nh}=0.
\end{equation}
Since at the event horizon, physics requires that the wave function can only propagate inward  and there can be no outward-propagating waves, we choose the inward-propagating solution
\begin{equation}
\phi_{\rm nh}(r) \sim \exp\left[ - i \left( \omega - \frac{q Q}{r_h} \right) r_* \right] 
\approx \exp\left[ - i \left( \omega - \frac{q Q}{r_h} \right) \left( \frac{1}{f'(r_h)} \ln|r - r_h| + C \right) \right],
\end{equation}
so that
\begin{equation}
\phi_{\rm nh}(r)=A_{\rm tra}|r - r_h|^{-i\kappa_H/\alpha}.\label{phi horizon}
\end{equation}
Here $A_{\rm tra}$ is a complex constant, and $\alpha=f'(r_h)=\frac{2M}{r_h^{2}}-\frac{2Q^{2}}{r_h^{3}}
+\frac{\lambda}{r_h^{2}}\Big(1-\ln\frac{r_h}{|\lambda|}\Big)$.

\subsection{Intermediate low-frequency solution}
In the overlap region II ($\omega M\ll1, mM\ll1, |qQ|\ll1$), we only consider $l = 0$ mode which is the dominant term in the absorption cross section computation. Thus the radial Eq.~(\ref {Near-horizon solution}) for this region reduces to
\begin{equation}
\frac{\textrm{d}^2}{\textrm{d}r^2}\phi_{\rm lf}+\left( \frac{2}{r} + \frac{f'(r)}{f(r)} \right) \frac{\textrm{d}}{\textrm{d}r}\phi_{\rm lf} = 0 ,
\end{equation}
with solution given by
\begin{equation}
\phi_{\rm lf}(r)=C_2 + C_1\int_{r_h}^r \frac{\textrm{d}r}{r^2 f(r)}\label{region2}.
\end{equation}
Consider behavior as $r\rightarrow r_h$,
\begin{equation}
\phi_{\rm lf}(r) \approx C_2+\frac{C_1}{\alpha}
\left[
\frac1{r_h^2}\ln\left|\frac{r-r_h}{r}\right|+\frac1{r_hr}
\right].\label{near region low frequency}
\end{equation}
Consider behavior as $r\to\infty$, 
\begin{equation}
\phi_{\rm lf}(r) \approx C_2'-\frac{C_1}{r}
+\mathcal O\!\left(\frac{\ln r}{r^2}\right)\label{far region low frequency}.
\end{equation}
The solutions near the horizon and in the far region represent the asymptotic behaviors of the same general solution in different regions, their constants are not independent but are related through a global integral:
\begin{equation}
C'_2 = C_2 + C_1 K_f.
\end{equation}
\begin{equation}
K_f= \int_{r_h}^{\infty}\left[\frac{1}{r^2 f(r)}-\frac{1}{\alpha r^2(r-r_h)}\right]dr.
\end{equation}
 We expand the regularized integral as $K_f=K_{\rm RN}+\lambda \delta K+O(\lambda^2)$, where 
zero-order term $K_{\rm RN}= \ln (r_+ - r_-)/(r_+ - r_-)$, $r_\pm = M \pm \sqrt{M^2 - Q^2}$, corresponding to the RN background.
First-order correction $\delta K$ arises from the \(\lambda\)-dependent term in \(f(r)\) and the shift of the event horizon position \(r_h\).
We may rewrite Eq.~(\ref{phi horizon}) as
\begin{equation}
\phi_{\rm nh}(r)=A_{\rm tra}(1-i\kappa_H\ln|r-r_h|/\alpha).\label{phi horizon2}
\end{equation}
Therefore, in the vicinity of the event horizon, the solution exhibits a logarithmic divergence behavior. Comparing the solutions in Region I Eq.~(\ref{phi horizon2}) and Region II  Eq.~(\ref{near region low frequency})  yields 
\begin{equation}
C_1=-ir_h^2 \kappa_H A_{\rm tra},\qquad C_2=A_{\rm tra},\qquad C'_2 = A_{\rm tra}  -ir_h^2 \kappa_H A_{\rm tra} K_f. \label{coefficient}
\end{equation}

\subsection{Far region Coulomb approximation}

In the far region III ($r\gg r_h$), the exact effective potential contains logarithmic long-range terms induced by $\lambda$. In the present approximation, these terms are treated as perturbative corrections, and the leading far-zone equation is approximated by the standard $l=0$ Coulomb equation, the equation of Eq.~(\ref{Near-horizon solution}) is
\begin{equation}
\left(\frac{\textrm{d}^2}{\textrm{d}r^2}+k^2+\frac{2k\eta_0}{r}\right)r\phi_{\rm far}=0,
\end{equation}
with
\begin{equation}
\eta_0=
-\frac{M(2\omega^2-m^2)-qQ\omega}{k}.
\end{equation}

The independent Coulomb solutions are $F_0(\eta_0,kr)$ and $G_0(\eta_0,kr)$, and we define the ingoing/outgoing combinations
\begin{equation}
H_0^{(\pm)}(\eta_0,kr)=G_0(\eta_0,kr)\pm iF_0(\eta_0,kr).
\end{equation}
The far-zone solution is thus written as
\begin{equation}
r\phi_{\rm far}=
A_{\mathrm{in}}\,H_0^{(-)}(\eta_0,kr)
+
A_{\mathrm{out}}\,H_0^{(+)}(\eta_0,kr).
\end{equation}
For $kr\ll 1$, the Coulomb wave functions behave as
\begin{equation}
F_0(\eta_0,kr)\sim \rho\,kr,
\qquad
G_0(\eta_0,kr)\sim \rho^{-1},
\end{equation}
where we have introduced the notation
\begin{equation}
\rho^2=\frac{2\pi\eta_0}{e^{2\pi\eta_0}-1}.
\end{equation}
Therefore,
\begin{equation}
H_0^{(\pm)}(\eta_0,kr)\sim \rho^{-1}\pm i\rho\,kr,
\end{equation}
and the far-region solution becomes
\begin{equation}
r\phi_{\rm far}
\approx
\rho^{-1}\left(A_{\mathrm{in}}+A_{\mathrm{out}}\right)
+i\rho k\left(A_{\mathrm{out}}-A_{\mathrm{in}}\right)r.\label{far region}
\end{equation}

\subsection{Absorption cross section}

Matching the coefficients of the linear and constant terms in far-region Eq.~(\ref{far region}) and over lap region Eq.~(\ref{far region low frequency}), we obtain
\begin{equation}
i\rho k\left(A_{\mathrm{out}}-A_{\mathrm{in}}\right)
=
A_{\rm tra}\left(1-i\kappa_H r_h^2 K_f\right),
\end{equation}
\begin{equation}
\rho^{-1}\left(A_{\mathrm{in}}+A_{\mathrm{out}}\right)
=
i\kappa_H r_h^2 A_{\rm tra}.
\end{equation}
Solving for $A_{\mathrm{out}}/A_{\mathrm{in}}$, the leading-order reflection coefficient is
\begin{equation}
R_{\omega l} \equiv \frac{A_{\mathrm{out}}}{A_{\mathrm{in}}}
=
\frac{
\left(1-i\kappa_H r_h^2 K_f\right)-\rho^2 k\kappa_H r_h^2
}{
\left(1-i\kappa_H r_h^2 K_f\right)+\rho^2 k\kappa_H r_h^2
}.
\end{equation}
At low frequencies, the dominant contribution comes from the $l=0$ mode
\begin{equation}
\sigma_{\mathrm{abs}}^{(0)}
=
\frac{\pi}{\omega^2 v^2}
\left(\frac{4\rho^2 \omega v \kappa_H r_h^2}
{\left(1+\rho^2 \omega v \kappa_H r_h^2\right)^2+\left(\kappa_H r_h^2 K_f\right)^2}\right).
\end{equation}
If we take $\lambda=0$, $q=0$, $K_f=\ln (r_+ - r_-)/(r_+ - r_-)$, it is the absorption cross section of the RN BH\cite{Benone2014}.

The effect of the parameter $\lambda$ enters the above formula implicitly through the horizon radius $r_h$ and the quantity $K_f$. The logarithmic long-range tail generated by $\lambda$ has been neglected at leading order in the far-zone wave equation and may be incorporated perturbatively as a higher-order phase correction.
In effective potential, we treat $\lambda$ term as a perturbation
\begin{equation}
\begin{aligned}
\delta V_\lambda(r)
&=-\frac{\lambda(2\omega^2-m^2)\ln(r/|\lambda|)}{r}
\\
&+\frac{
4\lambda(qQ\omega-3M\omega^2+Mm^2)
+\lambda^2(3\omega^2-m^2)(\ln(r/|\lambda|))^2}{r^2}
\end{aligned}
\end{equation}
It mainly modifies the phase
\begin{equation}
\Theta(r)=kr+\eta_0\ln(2kr)+\sigma_0+\delta\Theta_\lambda(r)
\end{equation}
where $\sigma_0 = \arg \Gamma(1+i\eta_0)$ with the leading correction 
\begin{equation}
\delta\Theta_\lambda(r)\approx
-\frac{\lambda(2\omega^2-m^2)}{4k}
\left(\ln\frac{r}{|\lambda|}\right)^2.
\end{equation}

\section{ High frequency regime}
\subsection{Geodesics and orbits}
In the gravitational field, charged massive particles subject to the Lorentz force imparted by the electric background field. The Lagrangian takes the form 
\begin{equation}
\mathcal{L}=\frac{1}{2} g_{\mu\nu}\dot{x}^{\mu} \dot{x}^{\nu}+\frac{q}{m}A_{\mu}\dot{x}^{\mu}\label{lag},
\end{equation}
where the overdot stands for the derivative with respect to the proper time, $q$ and $m$ are the charge and mass of the particle, respectively. Due to the assumption of stationarity and axisymmetry of the space-time, it has two  Killing vector $\partial/\partial t$ and $\partial/\partial \phi$ which in turn guarantees the existence of two conserved quantities for geodesic motion:
\begin{equation}
E=-m\frac{\partial \mathcal{L}}{\partial \dot{t}},\quad L=m\frac{\partial  \mathcal{L}}{\partial \dot{\phi}}\label{conserve}.
\end{equation}
Here, $E$ and $L$ are the energy and the angular momentum of the particle along a particle's orbit, respectively. The conserved quantities reduce the equations for $t$ and $\phi$ into first integrals:
\begin{equation}
\dot{t}= \frac{E + qA_0(r)}{mf(r)}, \quad \dot{\phi}= \frac{L}{mr^2}.
\end{equation}
In asymptotically flat spacetimes $ L$ and $E$ takes the form
\begin{equation}
L=\frac{mv}{\sqrt{1-v^2}},\quad E=\frac{m}{\sqrt{1-v^2}}\label{LE},
\end{equation}
where $v=\sqrt{1-m^2/\omega^2}$ is the asymptotic velocity, and the impact parameter is $b \equiv L/vE$.
We have the normalization condition $g_{\mu\nu} \dot{x}^{\mu}\dot{x}^{\nu}=\epsilon$ (where $\epsilon=-1, 0, 1$ correspond to time-like, null and spacelike geodesics respectively). The radial geodesics in the equatorial plane are governed by the orbital equation
\begin{equation}
 \dot{r}^2= \frac{[(E + qA_0(r))^2 - m^2f(r)]  r^2 - L^2f(r)}{m^2r^2}\equiv\Gamma(r)\label{geodesics}.
\end{equation}
From here we cannot directly obtain an effective potential similar to that of Schwarzschild spacetime. But we can determine the points where $\dot{r}=0$, represent turning points of the trajectory in which the particle reaches zero velocity call curves of zero velocity.

Perpetual circular orbits are possible if conditions $\dot{r}=0$ and $\ddot{r}=0$ can be satisfied simultaneously. We may find the critical radius $r_c$ of the unstable circular orbit and the critical impact parameter $b_c$, namely
\begin{equation}
\Gamma(r_c)=0, \quad \frac{\textrm{d}\Gamma(r)}{\textrm{d}r}\vline \underset{r=r_c} \vline =0\label{orbit1}.
\end{equation}
We may find the critical impact parameter $b_c$
\begin{equation}
b_c^2=\frac{r_c^2(1-v^2)[(m/\sqrt{1-v^2}+qA_0(r_c))^2-m^2 f(r_c)]}{m^2v^2 f(r_c)}\label{bc},
\end{equation}
and an equation determining the value of $r_c$
\begin{equation}
\begin{aligned}
&2f(r_c)\left(m^2(1-v^2)f(r_c)-(m+q\sqrt{1-v^2}A_0(r_c))[m+q\sqrt{1-v^2}(A_0(r_c)+r_cA_0^\prime(r_c))]\right)\\
&+r_c f^\prime(r_c)(m+q\sqrt{1-v^2}A_0(r_c))^2=0.
\end{aligned}
\label{rc_equation}
\end{equation}
In the limit $v\rightarrow1$ of the asymptotic velocity, we obtain
\begin{equation}
2f(r_c)-r_c f'(r_c)=0 \Longrightarrow 2r_c^2-(\lambda+6M)r_c+4Q^2+3\lambda r_c \textrm{ln} \frac{r_c}{|\lambda|}=0 \label{rc},
\end{equation}
and
\begin{equation}
b_c^2=\frac{r_c^2}{f(r_c)}\Longrightarrow b_c= \sqrt{\frac{3r_c^4}{r_c^2+\lambda r_c-Q^2 }}\label{bc2}.
\end{equation}
In this limit, $b_c$ is independent of the particle's mass $m$ and charge $q$, and is determined solely by the BH parameters $M$, $Q$, and $\lambda$. Particles coming from infinity with impact parameter $b_c$ orbit around the BH an infinite number of times at $r=r_c$. When $ b < b_c$ they have more energy than the barrier of the effective potential, will cross the event horizon, being absorbed by the BH. On the other hand, for $b > b_c$, particles with energy less than the barrier of the effective potential, will be scattered by the central object.

\begin{figure}[!t]
\centering
\begin{minipage}{0.5\textwidth}
    \centering
    \includegraphics[width=\linewidth, keepaspectratio]{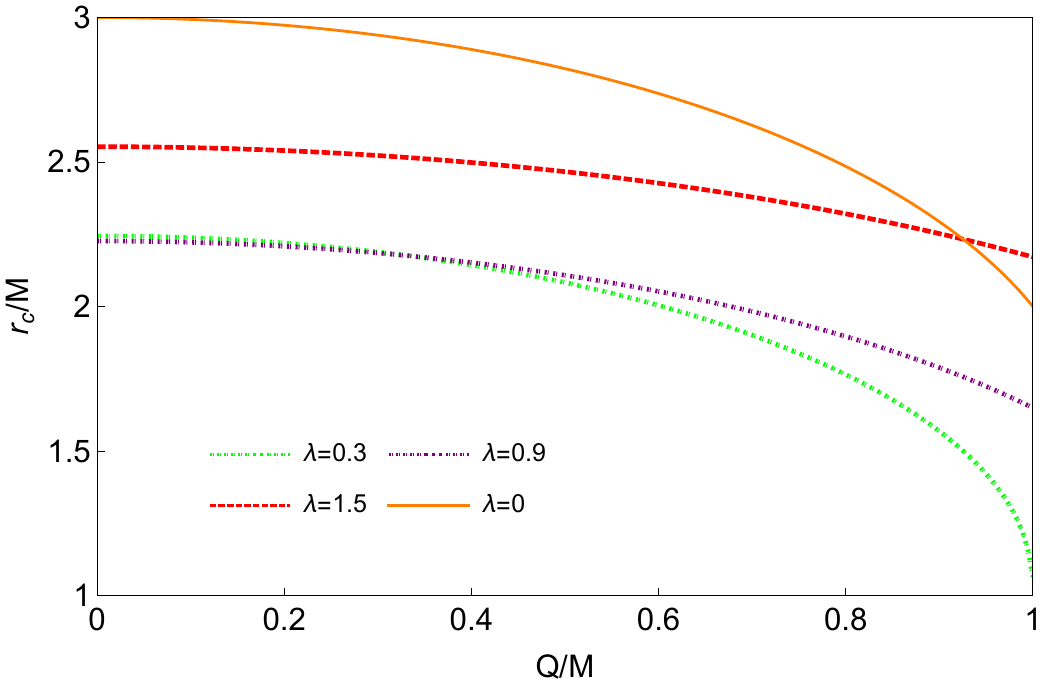}
\end{minipage}\hfill
\begin{minipage}{0.5\textwidth}
    \centering
    \includegraphics[width=\linewidth, keepaspectratio]{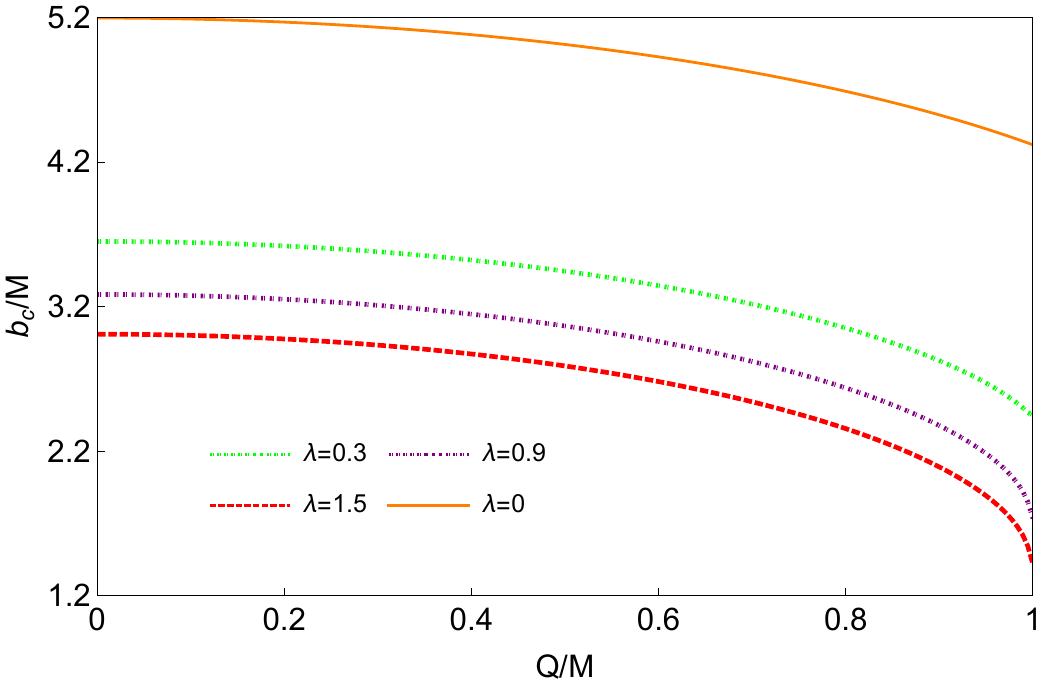}
\end{minipage}\hfill
\caption{The critical radius $r_c$ and critical impact parameter $b_c$ of PFDM and RN (solid lines, $\lambda=0$) BHs, as a function of $Q$, the plots are shown for $m=0$, $q=0$, $\lambda=0.3,0.9,1.5$.}
\label{rcbc1}
\end{figure}

\begin{figure}[!t]
\centering
\begin{minipage}{0.5\textwidth}
    \centering
    \includegraphics[width=\linewidth, keepaspectratio]{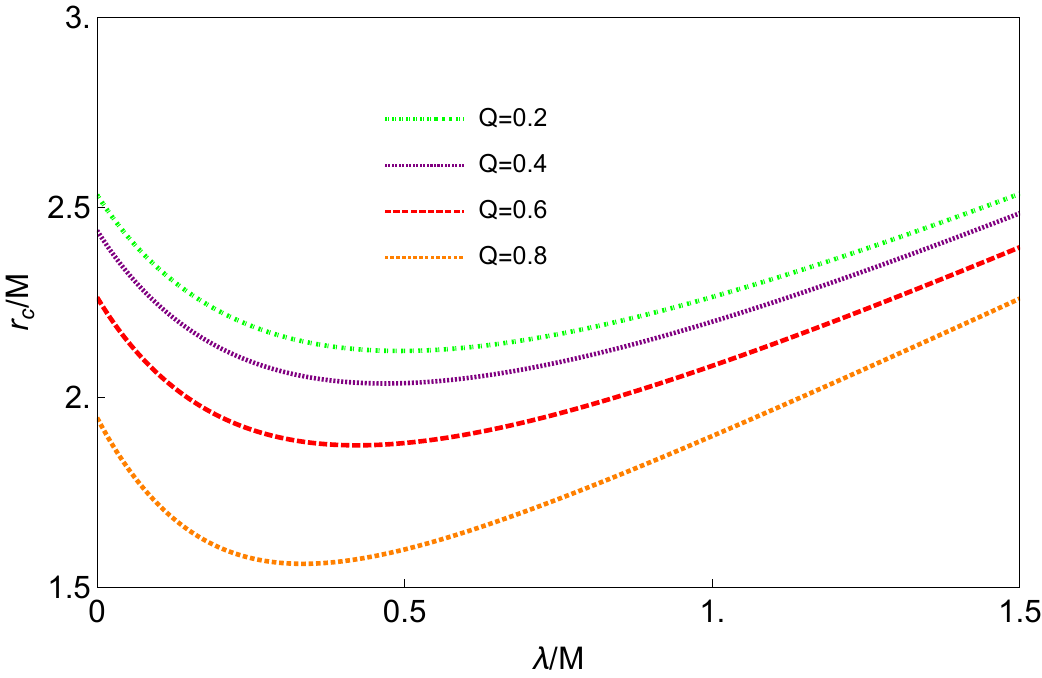}
\end{minipage}\hfill
\begin{minipage}{0.5\textwidth}
    \centering
    \includegraphics[width=\linewidth, keepaspectratio]{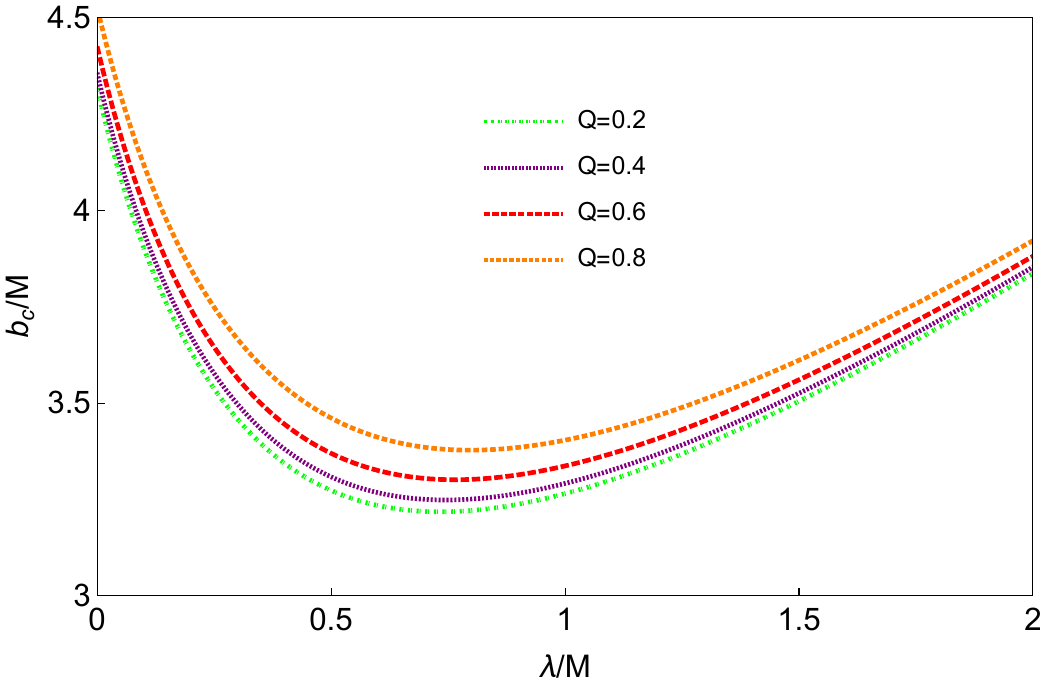}
\end{minipage}\hfill
\caption{The critical radius $r_c$ and critical impact parameter $b_c$ of PFDM BHs, as a function of  $\lambda$, the plots are shown for $m=0$, $q=0$, $Q=0.2,0.4,0.6,0.8$.}
\label{rcbc2}
\end{figure}

Eq.~(\ref{rc}) is a transcendental equation, when $Q=0$, we can find an analytical expression for the critical radius $r_c$ given by 
\begin{equation}
r_c=\frac{3}{2}\lambda \: \textrm{Lambert W}\left[\frac{2}{3}\exp\left(\frac{6M+\lambda}{3\lambda}\right)\right]\label{lambert}.
\end{equation}
The Lambert W function, denoted $W(z)$, is defined as the inverse function of $f(w)=we^w$, so $z=we^w$ implies $w=W(z)$. When $Q\neq 0$, by solving numerically we can compute $r_c$ for a given $Q$, $\lambda$ and consequently find $b_c$. In Fig.~\ref{rcbc1}, we show $r_c$ and $b_c$ for PFDM BHs as a function of $Q$. In comparison, the corresponding $b_c$ and $r_c$ for an RN BH with the same $Q$ is given. It is observed that $r_c$ and $b_c$ monotonically decrease as $Q$ increases while fixing $\lambda$ and larger $\lambda$ corresponding to smaller $b_c$. We note that, for a fixed value of $Q$, $b_c^{PFDM}<b_c^{RN}$, when $\lambda$ is small, $r_c^{PFDM}<r_c^{RN}$, while  $r_c^{PFDM}>r_c^{RN}$ for larger  $\lambda$ and $Q$. When $Q=0$, RN BHs tend to the Schwarzschild case, namely $r_c=3M$ and $b_c=3\sqrt{3}M$. In Fig.~\ref{rcbc2}, we show $r_c$ and $b_c$ for PFDM BHs as a function of $\lambda$. We find that for a fixed value of $\lambda$, $r_c$ decreases with an increase in charge $Q$ while $b_c$ exhibits the opposite behavior. Also, we found that  $r_c$ and $b_c$ initially decrease with an increase in $\lambda$ reaches a minimum in the interval $\lambda<\lambda_0$, and then starts to increase in the interval $\lambda>\lambda_0$, $\lambda=\lambda_0$ is the reflecting point. For fixed values of charge $Q$, Eq.~(\ref{rc}) is an implicit equations $f(r_c(\lambda),\lambda)=0$, at the reflecting point $r_c$ has a minimal value corresponding to the value of $\lambda_0$ meets conditions $\frac{d r_c}{d \lambda}=0$ and  $\frac{d^2 r_c}{d \lambda ^2}>0$. With this condition, we can get
\begin{equation}
\frac{\partial f}{\partial r_c}\frac{\textrm{d} r_c}{\textrm{d} \lambda}+\frac{\partial f}{\partial \lambda}=0  \Longrightarrow (r_c)_{min}=\lambda _0 e^{\frac{4}{3}}\label{implicit}.
\end{equation}
Here $e$ is the base of the natural logarithm, using the expression for the critical radius (\ref{rc}), we can determine the reflection point
\begin{equation}
\lambda_0=\frac{3M+\sqrt{9M^2-4Q^2\left(2+3e^{-4/3}\right)}}{2e^{4/3}+3}\label{implicit2}.
\end{equation}

For a BH having a photon sphere, classical capture cross section of geodesics coincides with the geometrical cross section of the photon sphere $\sigma_{geo} \equiv\pi b_c^2$. On the other hand, $r_c$ represent the circular orbit of the photon, that is, the photon sphere radius. In the spherically symmetric scenario, the shadow radius $r_{sh}$ turns out to be equal to $b_c$  as seen by 
an observer located at infinity\cite{Paula2023}.

\subsection{Classical scattering}
Defining $u = 1/r$, the orbit equation for charged massive particles is
\begin{equation}
\left(\frac{\textrm{d}u}{\textrm{d}\varphi}\right)^2=\frac{(1+\frac{\sqrt{1-v^2}}{m}qA_0(u))^2}{b^2 v^2}-f(u) \left(\frac{1-v^2}{b^2v^2}+u^2\right)   \equiv \mathcal{T}(u)\label{jiaodu}.
\end{equation}
The deflection angle of the scattered massive charged particle is given by
\begin{equation}
\Theta(b)=2\int_0^{u_0}\frac{\textrm{d}u}{\sqrt{\mathcal{T}}}-\pi\label{angle},
\end{equation}
here $u_0=1/r_0$, where $r_0$ is the distance of closest approach and satisfies $\mathcal{T}(u)|_{u=u_0}=0$. 
To calculate the deflection angle analytically, we restrict ourselves to the weak-field limit, where the impact parameter is much larger than the characteristic length scales of the system, that is $Mu \ll 1$, $Qu \ll 1 $, $\lambda u\left|\ln\frac{1}{u\lambda}\right|\ll1$. Including contributions up to second order, the weak-field deflection angle is given by (See Appendix A )

\begin{equation}
\begin{aligned}
\Theta(b)
&\approx\frac1b\Bigg[
2M\left(1+\frac1{v^2}\right)
-\frac{2qQ\sqrt{1-v^2}}{mv^2}
-\lambda\left(
\left(1+\frac1{v^2}\right)\ln\frac{b}{2|\lambda|}
+1
\right)
\Bigg]
\\[4pt]
&+\frac{\pi}{b^2}\Bigg[
\frac{3(4+v^2)}{4v^2}M^2
-\frac{3MqQ\sqrt{1-v^2}}{mv^2}
+\frac{q^2Q^2(1-v^2)}{2m^2v^2}
-\frac{2+v^2}{4v^2}Q^2
\\[4pt]
&+M\lambda\left(
\frac{3}{8}+\frac{3}{v^2}+\frac{1}{2v^4}
-\frac{3(4+v^2)}{4v^2}\ln\frac{2b}{|\lambda|}
\right)-\frac{qQ\sqrt{1-v^2}}{mv^2}\lambda
\left(
\frac{1-v^2}{2v^2}+2-\frac{3}{2}\ln\frac{b}{2|\lambda|}
\right)
\\[4pt]
&+\lambda^2\left(
\frac{3(4+v^2)}{16v^2}\ln^2\frac{2b}{|\lambda|}
-\frac{1}{16}\left(3+\frac{24}{v^2}+\frac{4}{v^4}\right)\ln\frac{2b}{|\lambda|}
+\frac{1-v^4}{4v^4}
+\frac{4+v^2}{64v^2}\pi^2
+\frac{5}{32}
\right)
\Bigg].
\end{aligned}
\label{scattering second order}
\end{equation}
The classical differential scattering cross section is given by \cite{Macedo2016}
\begin{equation}
\frac{\textrm{d}\sigma}{\textrm{d}\Omega}=\sum_i \frac{b_i}{\sin \theta}\left|\frac{\textrm{d}b_i}{\textrm{d}\theta}\right|\label{Classical scattering}.
\end{equation}
By using Eq.~(\ref{scattering second order}), 
define
\begin{equation}
\beta \equiv 1+\frac{1}{v^2}, 
\qquad 
\varepsilon \equiv \frac{qQ\sqrt{1-v^2}}{m v^2},
\qquad
K \equiv 2M\beta-2\varepsilon.
\end{equation}
Keeping the leading small angle contribution together with the first-order correction in $\lambda$, one obtains (See Appendix B )
\begin{equation}
\frac{\textrm{d}\sigma}{\textrm{d}\Omega}
\approx
\frac{1}{\theta^4}
\left[
K^2
-\lambda K\left(
2\beta\ln\frac{|K|}{2|\lambda| \theta}
+\beta+2
\right)
\right].\label{weak field expression}
\end{equation}
When $\lambda=0$, the leading term corresponds to the generalized Rutherford formula, which includes the combined contributions from gravitational and electromagnetic interactions.
\begin{equation}
\begin{aligned}
\frac{\textrm{d}\sigma}{\textrm{d}\Omega}
&  \approx
\frac{1}{\theta^4}
\left[
2M\left(1+\frac{1}{v^2}\right)-\frac{2qQ\sqrt{1-v^2}}{m v^2}
\right]^2
\\
& +\frac{\pi}{\theta^3}
\left[
\frac{3(4+v^2)}{4v^2}M^2
-\frac{3M qQ\sqrt{1-v^2}}{m v^2}
+\frac{q^2Q^2(1-v^2)}{2m^2v^2}
-\frac{2+v^2}{4v^2}Q^2
\right].\label{weak field expression2}
\end{aligned}
\end{equation}

\begin{figure}[!t]
\centering
\begin{minipage}{0.5\textwidth}
    \centering
    {\fontsize{8pt}{10pt}\selectfont $q=0.3$, $\lambda=0.1$}\\
    \includegraphics[width=\linewidth, keepaspectratio]{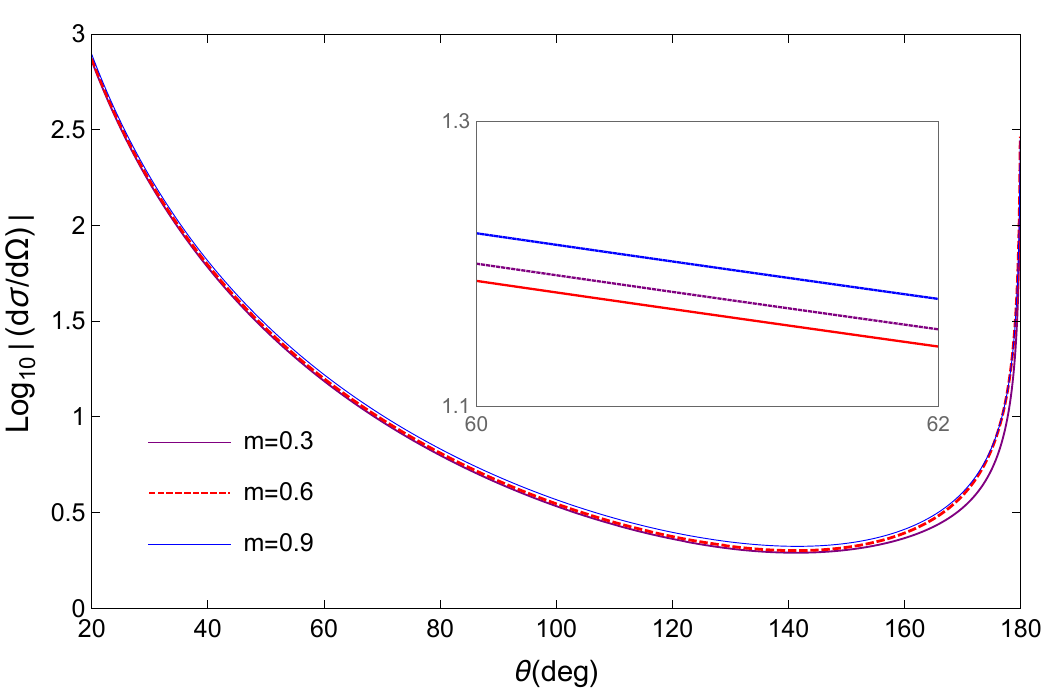}
\end{minipage}\hfill
\begin{minipage}{0.5\textwidth}
    \centering
    {\fontsize{8pt}{10pt}\selectfont $q=0.3$, $\lambda=0.3$}\\
    \includegraphics[width=\linewidth, keepaspectratio]{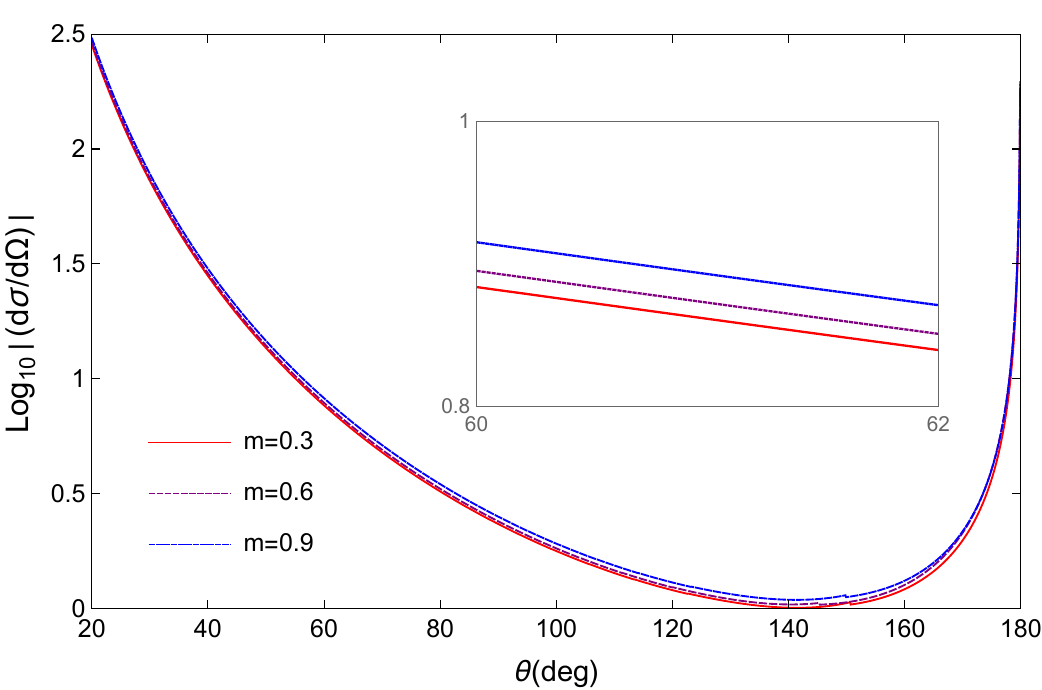}
\end{minipage}\hfill
\vspace{0.2cm}
\begin{minipage}{0.5\textwidth}
    \centering
    {\fontsize{8pt}{10pt}\selectfont $m=0.3$, $\lambda=0.1$}\\
    \includegraphics[width=\linewidth, keepaspectratio]{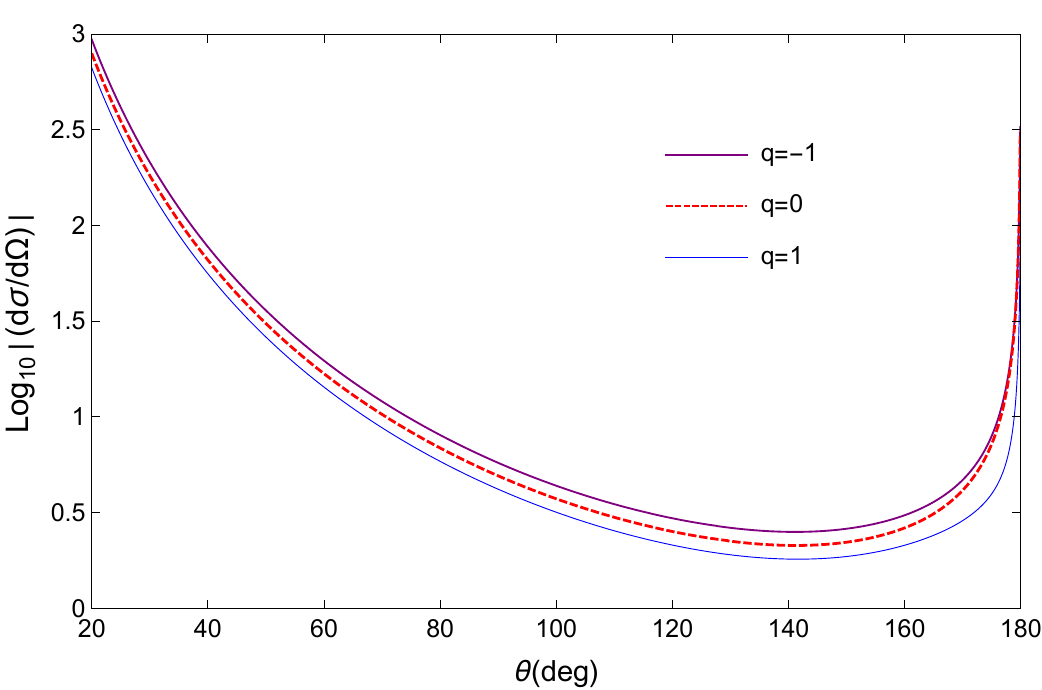}
\end{minipage}\hfill
\begin{minipage}{0.5\textwidth}
    \centering
    {\fontsize{8pt}{10pt}\selectfont $m=0.3$, $\lambda=0.3$}\\
    \includegraphics[width=\linewidth, keepaspectratio]{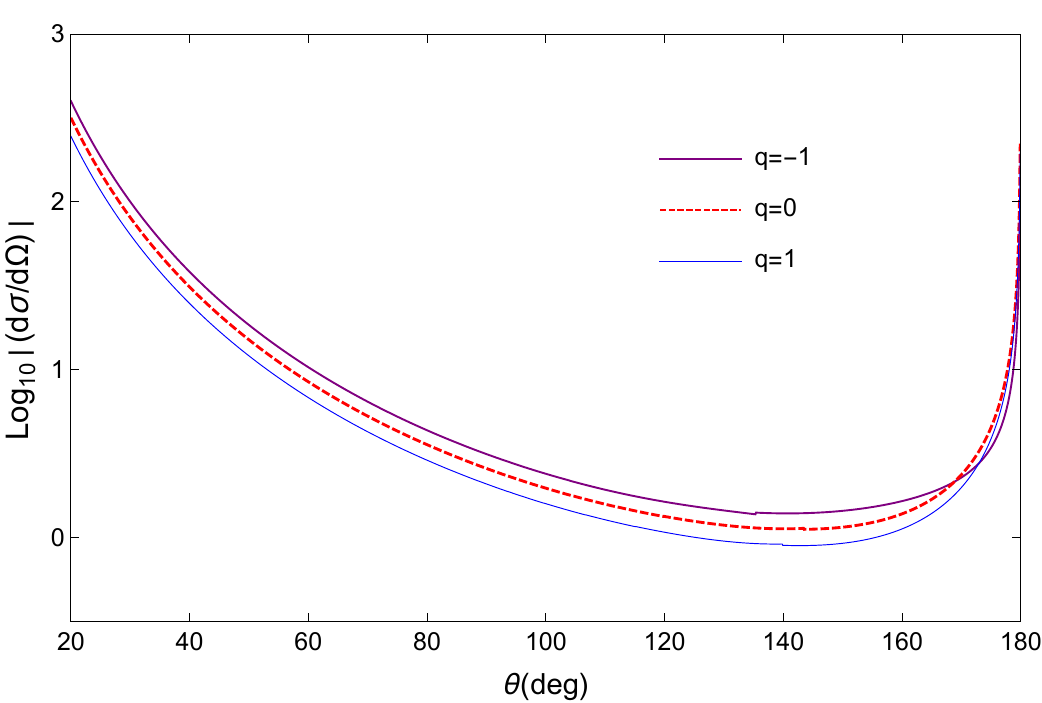}
\end{minipage}\hfill
\vspace{0.2cm}
\begin{minipage}{0.5\textwidth}
    \centering
    {\fontsize{8pt}{10pt}\selectfont $m=0.3$, $q=0.3$}\\
    \includegraphics[width=\linewidth, keepaspectratio]{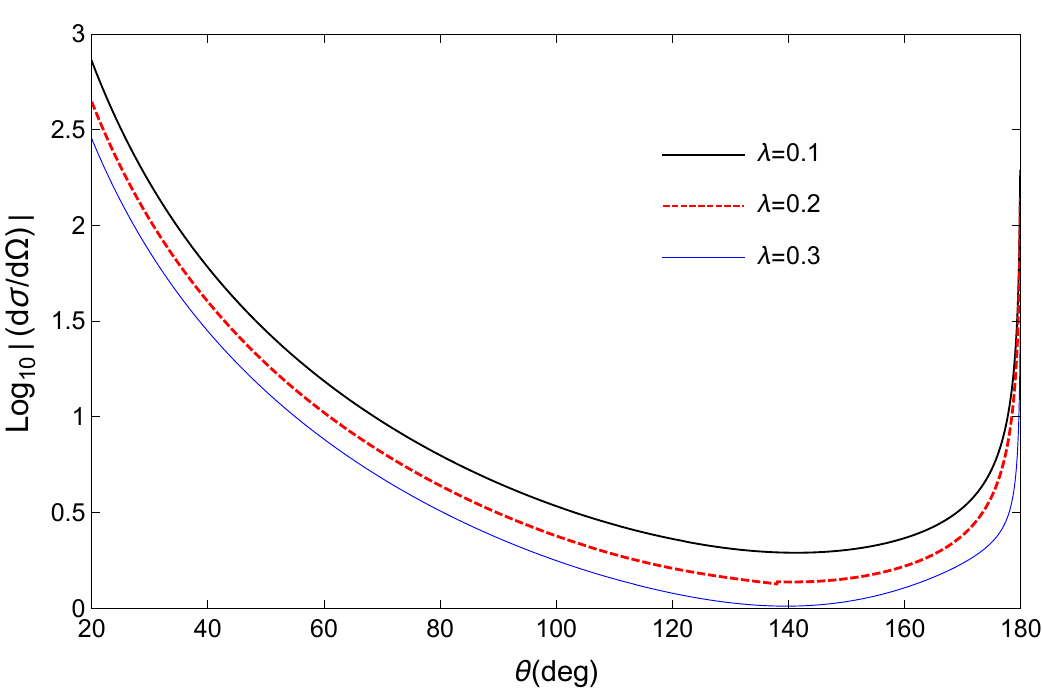}
\end{minipage}\hfill
\begin{minipage}{0.5\textwidth}
    \centering
    {\fontsize{8pt}{10pt}\selectfont $m=0.3$, $q=0.3$}\\
    \includegraphics[width=\linewidth, keepaspectratio]{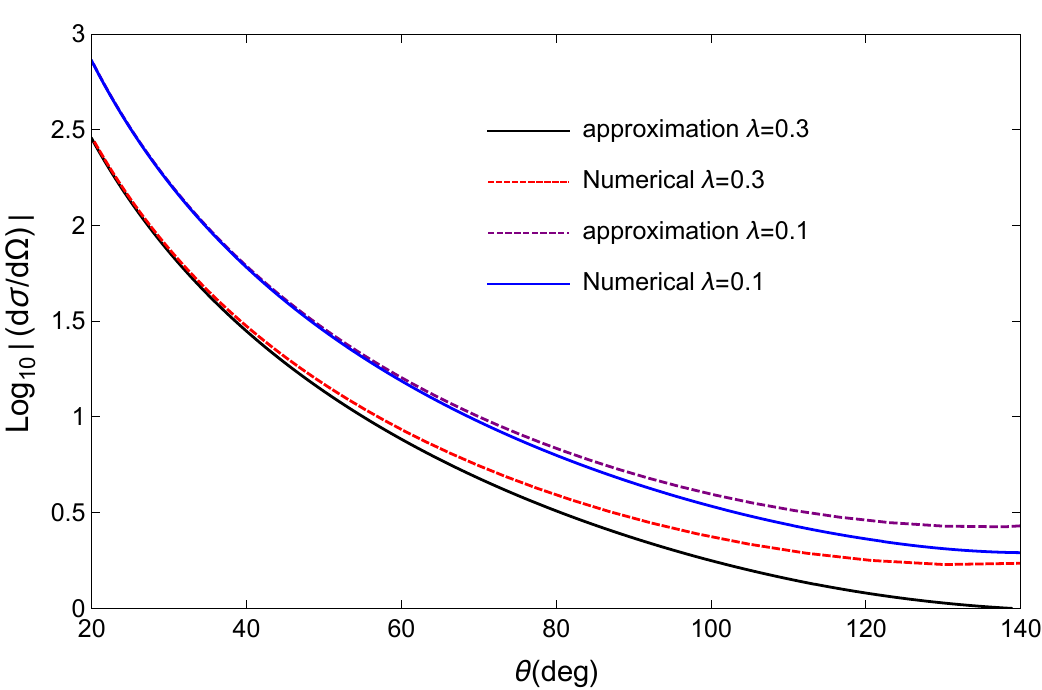}
\end{minipage}\hfill
\caption{\label{classical scattering} The classical scattering cross section of RN BH immersed in PFDM for different values of $\lambda$, $q$ and $m$. We set $M=1$, $Q=0.5$, $\omega=3$. Under the same parameters, comparison between the exact numerical calculation and weak-field approximation results (bottom-right). 
}
\end{figure}

In the subsequent calculations, we set $M$ to unity, with all mass-dimension quantities rescaled by $M$, including $Q$, $q$, $m$ and $\lambda$, as shown in the plots. In Fig.~\ref{classical scattering}, we have investigated the influence of different parameters $\lambda$, $q$, $m$ on the classical differential scattering cross section $\mathrm{d}\sigma/\mathrm{d}\Omega$, two representative cases with $\lambda=0.1$ and $\lambda=0.3$ are considered. For all parameter choices, $\mathrm{d}\sigma/\mathrm{d}\Omega$ decreases monotonically with increasing scattering angle $\theta$. In logarithmic scale, $\log_{10}|\mathrm{d}\sigma/\mathrm{d}\Omega|$ exhibits a smooth decay from small to large angles, indicating that forward scattering dominates the process. Case I (top-left, top-right): At $Q=0.5$, $q=0.3$, $\omega=3$ with $\lambda=0.1$ and $\lambda=0.3$, increasing the particle mass $m$ leads to a systematic enhancement of $\mathrm{d}\sigma/\mathrm{d}\Omega$ over the entire angular range, it does not change the shape (the angular dependence remains almost unchanged), but only modifies the amplitude. Case II (center-left, center-right): At $Q=0.5$, $m=0.3$, $\omega=3$ with $\lambda=0.1$ and $\lambda=0.3$, increasing the particle charge $q$, from $-1$ to $1$, leads to a decrease in the amplitude of $\mathrm{d}\sigma/\mathrm{d}\Omega$ over the entire angular range. Neither the particle mass $m$ nor the charge $q$ significantly alters the overall angular dependence of the differential scattering cross section. We find that, as $\lambda$ increases, the scattering amplitude is significantly reduced (bottom-left). Comparing the two cases of $m$ and $q$, the overall magnitude of the cross section is reduced for high $\lambda$ value. These results suggest that the logarithmic dark matter term tends to weaken both the overall scattering strength and its dependence on $m$ and $q$.

We also show the comparison of the classical scattering cross section between the weak-field analytical Eq.~(\ref{weak field expression}) and numerical results using Eq.~(\ref{angle}) for $\lambda=0.1, 0.3$. The results are in excellent agreement in the forward direction. The result for $\lambda=0.3$ deviates more rapidly from the numerical calculation as the scattering angle increases, whereas the smaller value $\lambda=0.1$ yields better agreement.

\subsection{Glory scattering}
In physics, glory scattering is a special scattering phenomenon, characterized by an anomalous enhancement of the scattering intensity directly in the backward direction of the incident beam. The semiclassical approximation~\cite{Matzner1985} used to describe this effect is valid in the high-energy or short-wavelength limit, although it fails to describe the differential scattering cross section at low angles:
\begin{equation}
\frac{\textrm{d}\sigma}{\textrm{d}\Omega}
\simeq
2\pi \omega v b_g^2
\left|\frac{\textrm{d}b}{\textrm{d}\theta}\right|_{\theta=\pi}
J_0^2\left(\omega v b_g\sin\theta\right)\label{Glory scattering},
\end{equation}
where $b_g$ is the impact parameter for backscattered waves, and $J_0$ is the zeroth-order Bessel function of the first kind. The semiclassical model of glory scattering describes wave interference effects. In the backward direction, the scattering angle is insensitive to variations in the impact parameter, leading to concentrated contributions from a large number of classical trajectories \cite{Macedo2016}. Oscillatory interference patterns arise from the phase differences between these trajectories.

\section{Numerical analysis and comparisons}

In this section, we discuss the absorption and scattering cross sections obtained by solving the radial equation Eq.~(\ref{radial}), numerically. To calculate the scattering amplitude we must first determine the phase shifts, we straightforwardly integrate the radial equation  from the event horizon to a large distance, we actually start integrating from a point outside the horizon that is very close to the horizon with
\begin{equation}
\psi(r)=
e^{-i\kappa_H r_*}
\sum_{n=0}^\infty a_n (r-r_h)^n.
\end{equation}
We determine the ingoing and outgoing coefficients by comparing the numerical solutions with the asymptotic forms in Eq.~(\ref{boundary}), and then we obtain the reflection and transmission coefficients.

\subsection{Absorption cross section}

\begin{figure}[!t]
\centering
\begin{minipage}{0.5\textwidth}
    \centering
    {\fontsize{8pt}{10pt}\selectfont $Q=0.5$, $q=0$, $m=0$}\\
    \includegraphics[width=\linewidth, keepaspectratio]{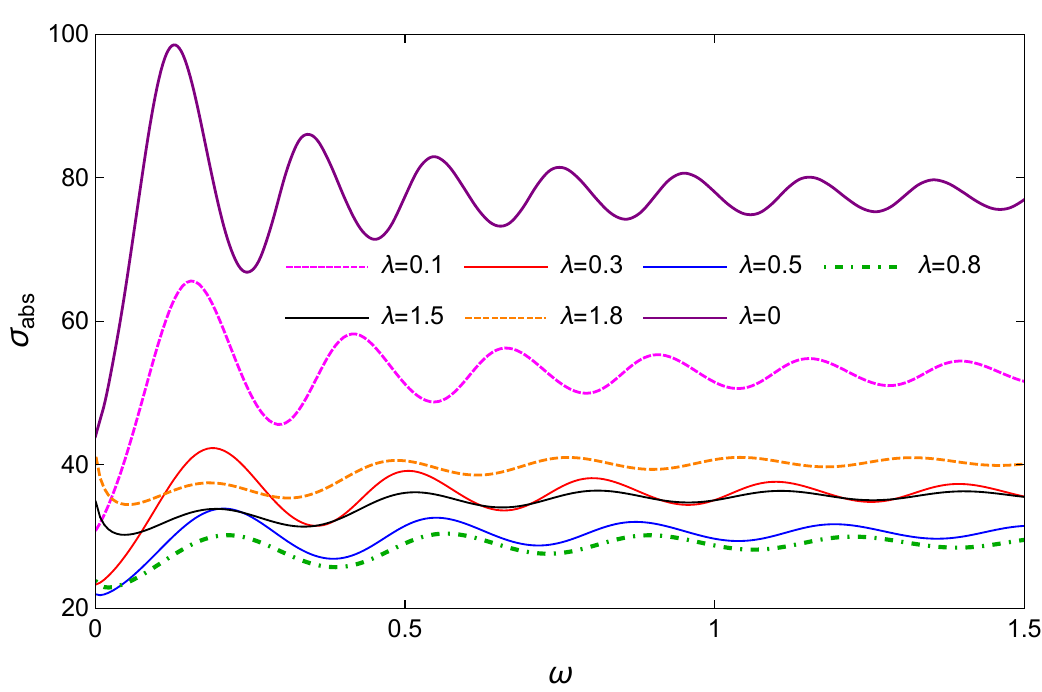}
\end{minipage}
\caption{\label{lambda} Total absorption cross sections for $\lambda=0, 0.1, 0.3, 0.5, 0.8, 1.5, 1.8$. We set $M=1$, $Q=0.5$, $m=0$, $q=0$.}
\end{figure}
Fig.~\ref{lambda} shows the absorption cross sections $\sigma_{abs}$ for $Q = 0.5$, $m=0$, $q=0$, and different choices of $\lambda$. When $\lambda=0$, the curve is obviously higher than the others across the whole frequency range, especially at low frequencies. When $\lambda\neq 0$, by contrast, the curve are suppressed to a narrow range, with the increase of \(\lambda\), the absorption cross section declines overall, and gradually increases again after attaining the critical value \(\lambda_0\). For any value of $\lambda$, the curves exhibit periodic oscillations as the frequency increases, indicating partial-wave interference structure of the system does not disappear with the variation of $\lambda$. However, as $\lambda$ increases, the difference between the peaks and valleys of the oscillations gradually decreases, and the curves become smoother. 

\begin{figure}[!t]
\centering
\begin{minipage}{0.5\textwidth}
    \centering
    {\fontsize{8pt}{10pt}\selectfont $Q=0.5$, $q=0$, $m=0$}\\
    \includegraphics[width=\linewidth, keepaspectratio]{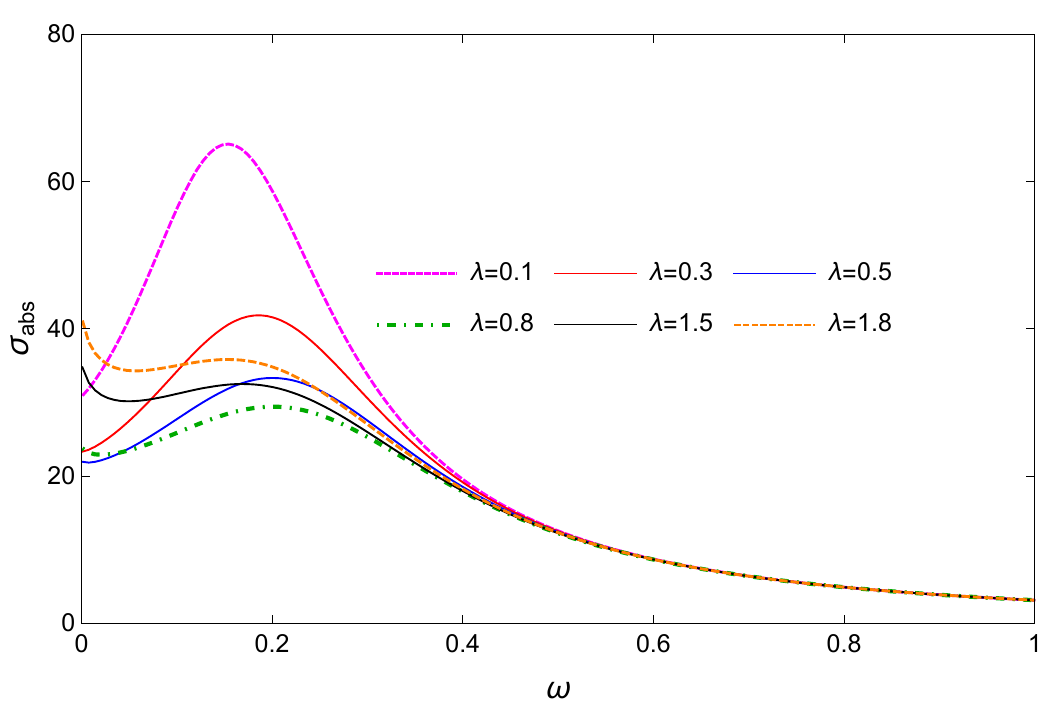}
\end{minipage}
\caption{\label{lambdal=0} : The partial absorption cross section with $l=0$ is plotted for PFDM BHs with $\lambda = 0, 0.1, 0.3, 0.5, 0.8, 1.5, 1.8$, where we set $M = 1$, $Q = 0.5$, $m = 0$, $q = 0$.}
\end{figure}
Here we plot the partial-wave  $\sigma_{abs}$ for $l=0$ corresponding to different values of $\lambda$ in Fig.~\ref{lambdal=0}. It can be seen that the peak of the partial wave gradually decreases as $\lambda$ increases, and a local minimum and a local maximum emerge after $\lambda$ exceeds a certain value. When $\omega\rightarrow 0$, the $\sigma_{abs}$ tend to the horizon area $4\pi r_h^2$.

\begin{figure}[!t]
\centering
\begin{minipage}{0.5\textwidth}
    \centering
    {\fontsize{8pt}{10pt}\selectfont $Q=0.4$, $q=0$, $m=0.3$}\\
    \includegraphics[width=\linewidth, keepaspectratio]{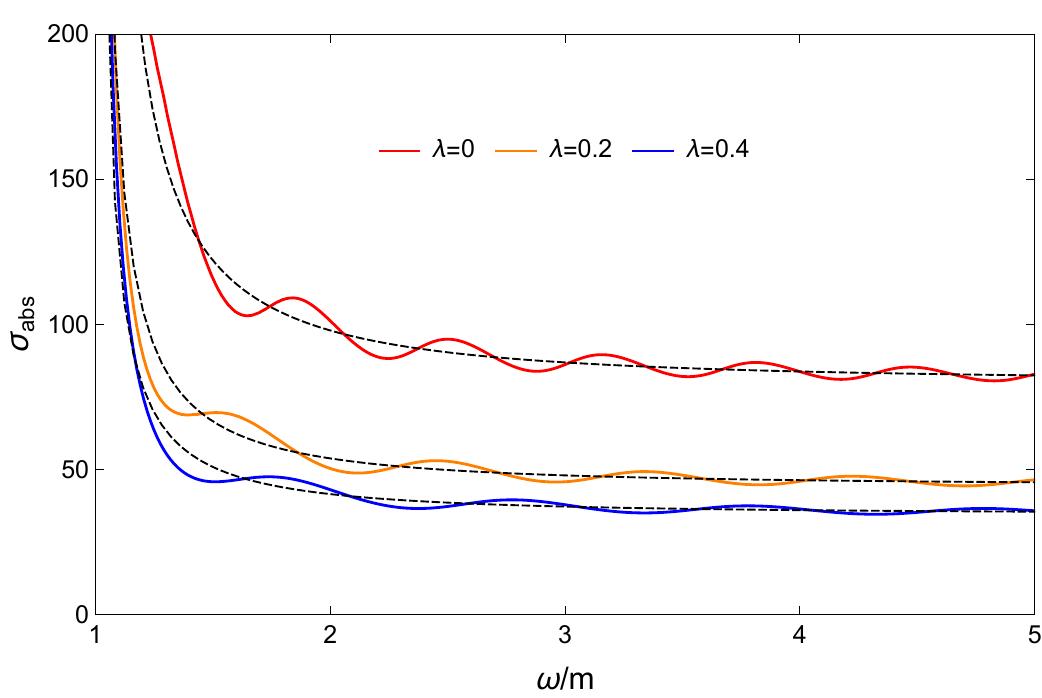}
\end{minipage}
\caption{\label{absorption m lambda} Total absorption cross sections of massive scalar waves by PFDM BHs for $M=1$, $Q=0.4$, $q=0$, $m=0.3$ and different choices of $\lambda$, grey dashed lines corresponding to $\pi b_c^2$.}
\end{figure}
In Fig.~\ref{absorption m lambda}, we show the influence of different $\lambda$ values on the $\sigma_{abs}$ at the same mass $m=0.3$. It can be seen that $\sigma_{abs}$ diminishes as $\lambda$ increases, and $\sigma_{abs}$ exhibits regular oscillations around $\pi b_c^2$. At low frequencies, the $\sigma_{abs}$ tend to diverge, no longer possess a finite $\sigma_{abs}$ like massless particles; In the high-frequency limit ($\omega/m\gg1$), for different values of $\lambda$, the $\sigma_{abs}$ converges to different asymptotic limits.

\begin{figure}[!t]
\centering
\begin{minipage}{0.5\textwidth}
    \centering
    {\fontsize{8pt}{10pt}\selectfont $Q=0.4$, $M=1$, $m = 0.4$, $q=1.2$}\\
    \includegraphics[width=\linewidth, keepaspectratio]{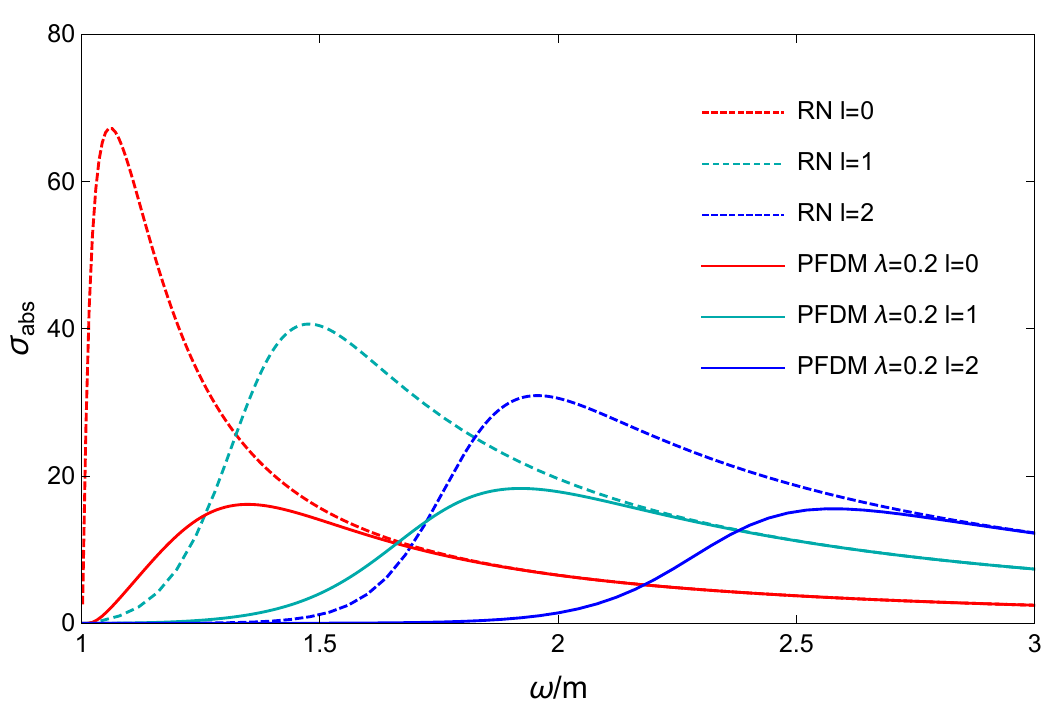}
\end{minipage}
\caption{\label{different_lambda_l=012} Partial absorption cross sections of charged massive scalar waves by RN and PFDM BHs for $l=0$ to $l=2$. }
\end{figure}
In Fig.~\ref{different_lambda_l=012}, we compare partial $\sigma_{abs}$ by RN and PFDM BHs, the $\sigma_{abs}$ of each partial wave for the PFDM BH is suppressed, especially for the \(l=0\) partial wave. The suppression occurs in the low-frequency region of the waveform for each partial wave, while in the high-frequency region, the waveform approaches that of the RN BH. Thus, the frequency corresponding to the peak of the same $l$ partial wave shifts toward higher frequencies.

\begin{figure}[!t]
\centering
\begin{minipage}{0.5\textwidth}
    \centering
    {\fontsize{8pt}{10pt}\selectfont $Q=0.4$, $M=1$, $m=0.4$, $\lambda=0.2$}\\
    \includegraphics[width=\linewidth, keepaspectratio]{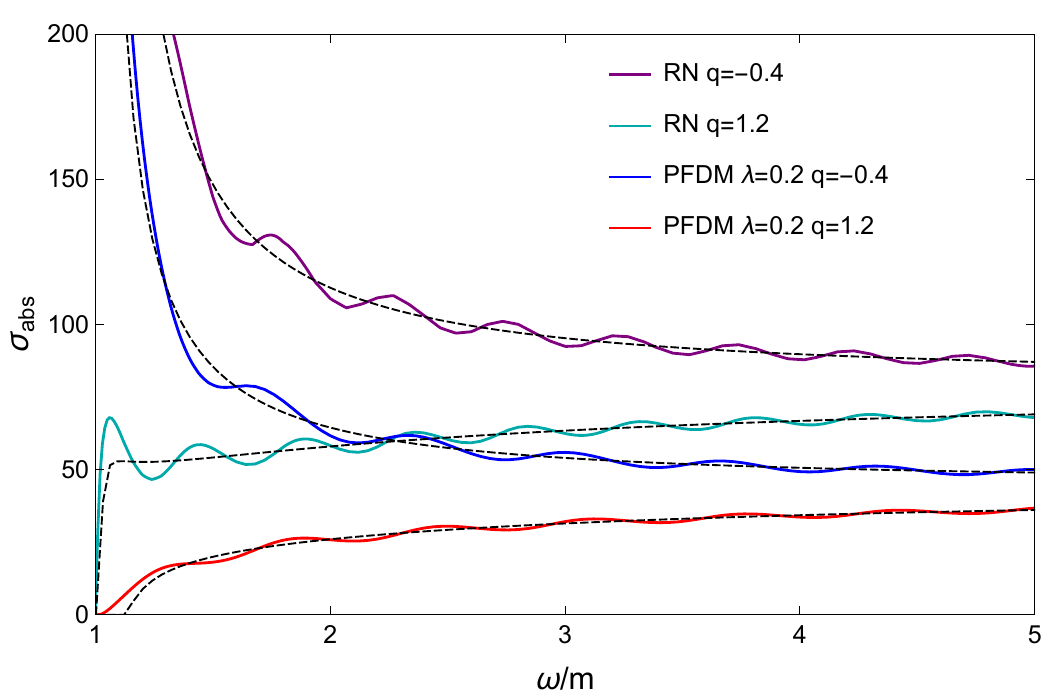}
\end{minipage}
\caption{\label{absorption cross section q m lambda} Total absorption cross sections of charged massive scalar waves by RN and PFDM BHs for different charge $q$. Grey dashed lines correspond to $\pi b_c^2$.}
\end{figure}
In Fig.~\ref{absorption cross section q m lambda}, we plot the total $\sigma_{abs}$ of RN and PFDM BHs with different charges. When a massive charged particle moves in this background, it is subjected to not only gravitational interaction but also electromagnetic force. It can be seen that $\sigma_{abs}$ is greatly suppressed at both charges $q=-0.4$ and $q=1.2$. In the $q=-0.4$ case, both $\sigma_{abs}$ tend to infinity when $\omega/m \rightarrow 1$, while in $q=1.2$ case, both $\sigma_{abs}$ tends to zero. In the case of $q=-0.4$, as $\omega \rightarrow \infty$, the RN BH and the PFDM BH exhibit different values of classical capture cross section $\pi b_c^2$. In the case of $q=1.2$, as $\omega \rightarrow \infty$, the RN BH and the PFDM BH also exhibit different values of $\pi b_c^2$. We can see that $\pi b_c^2$ in the case of RN $q=-0.4$ and PFDM $\lambda=0.2$, $q=-0.4$ tend to be consistent. Similarly, $\pi b_c^2$ in the case of RN $q=1.2$ and PFDM $\lambda=0.2$, $q=1.2$ also tend to be consistent. From Eq.~(\ref{rc}) and Eq.~(\ref{bc2}), it is clear that the mass $m$ and charge $q$ no longer affect the dominant result of $\pi b_c^2$ in the high-frequency limit; instead, the result depends on the values of $M$, $Q$ and $\lambda$. 

\begin{figure}[!t]
\centering
\begin{minipage}{0.5\textwidth}
    \centering
    {\fontsize{8pt}{10pt}\selectfont $Q=0.7$, $q=1.2$, $m=0.4$, $\lambda=0.2$}\\
    \includegraphics[width=\linewidth, keepaspectratio]{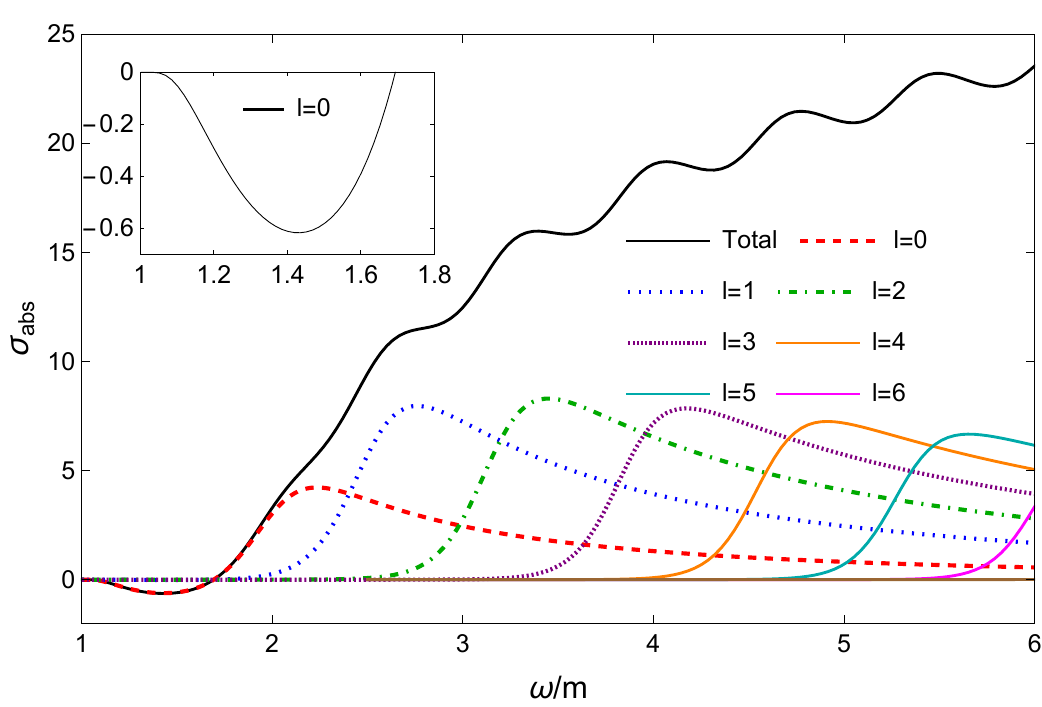}
\end{minipage}\hfill
\begin{minipage}{0.5\textwidth}
    \centering
    {\fontsize{8pt}{10pt}\selectfont $Q=0.7$, $q=1.2$, $m=0.4$}\\
    \includegraphics[width=\linewidth, keepaspectratio]{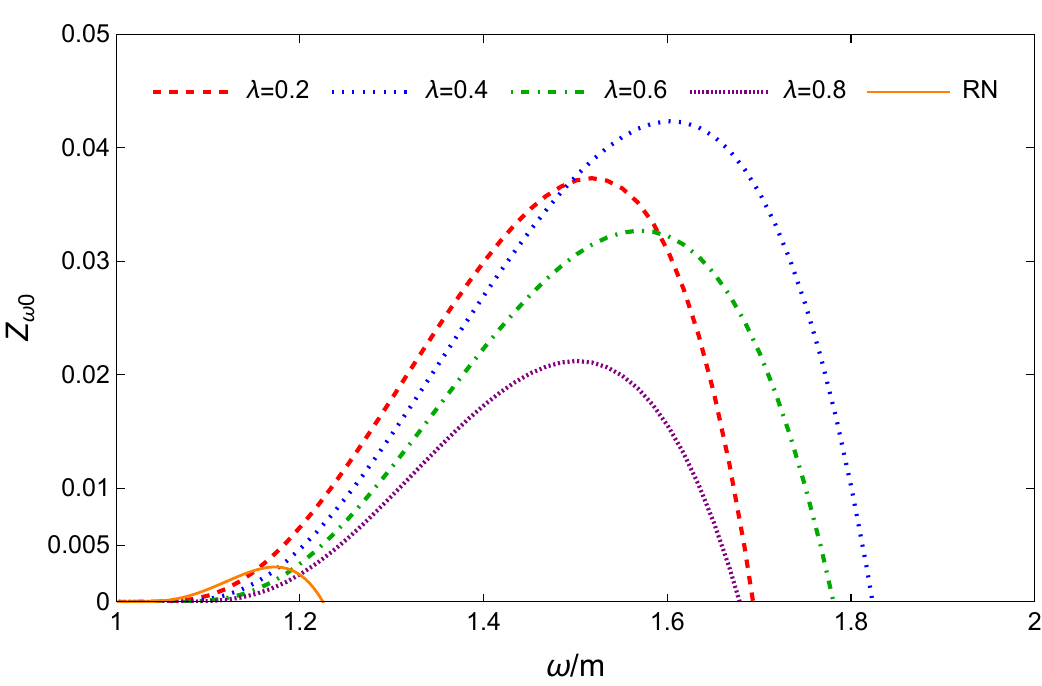}
\end{minipage}\hfill
\caption{Partial and total absorption cross sections of charged massive scalar fields by PFDM BHs (left); Superradiant amplification factor of charged massive scalar fields by PFDM and RN BHs for $\lambda=0.2,0.4,0.6,0.8$, $l=0$ (right).}
\label{superradiance}
\end{figure}
In Fig.~\ref{superradiance}, we present the partial and total $\sigma_{abs}$ for $Q=0.7$, $q=1.2$, $m=0.4$, $\lambda=0.2$ (Left figure). For frequencies $m<\omega<qQ/r_h$, this implies superradiance, and the outgoing wave will be amplified at infinity. As can be seen from the figure, $\sigma_{abs}$ becomes negative when $\omega$ is approximately less than $1.7m$. The dominant contribution to superradiance comes from the monopole mode $l=0$, the $l=1$ partial wave is only of an extremely small order of magnitude. For the region where the $\sigma_{abs}$ is positive, the contribution of the $l=0$ partial wave is instead smaller than that of the $l=1$ partial wave. A comparison of the amplification factors $Z_{\omega l}=|R_{\omega l}|^2-1$ in the PFDM and RN geometries is also exhibited (right figure). For superradiance, the amplification factor becomes positive. The superradiant amplification in the background of the PFDM BH, for fixed values of the BH parameters $Q$, $q$, and $m$, is typically larger than that in the corresponding RN geometry. As $\lambda$ increases, the amplification factor rises to a critical value and then decreases with further growth of $\lambda$.

\subsection{Differential scattering cross section}

\begin{figure}[!t]
\centering
\begin{minipage}{0.5\textwidth}
    \centering
    {\fontsize{8pt}{10pt}\selectfont $q=0.2$,$m=0.3$,$\lambda=0.05$}\\
    \includegraphics[width=\linewidth, keepaspectratio]{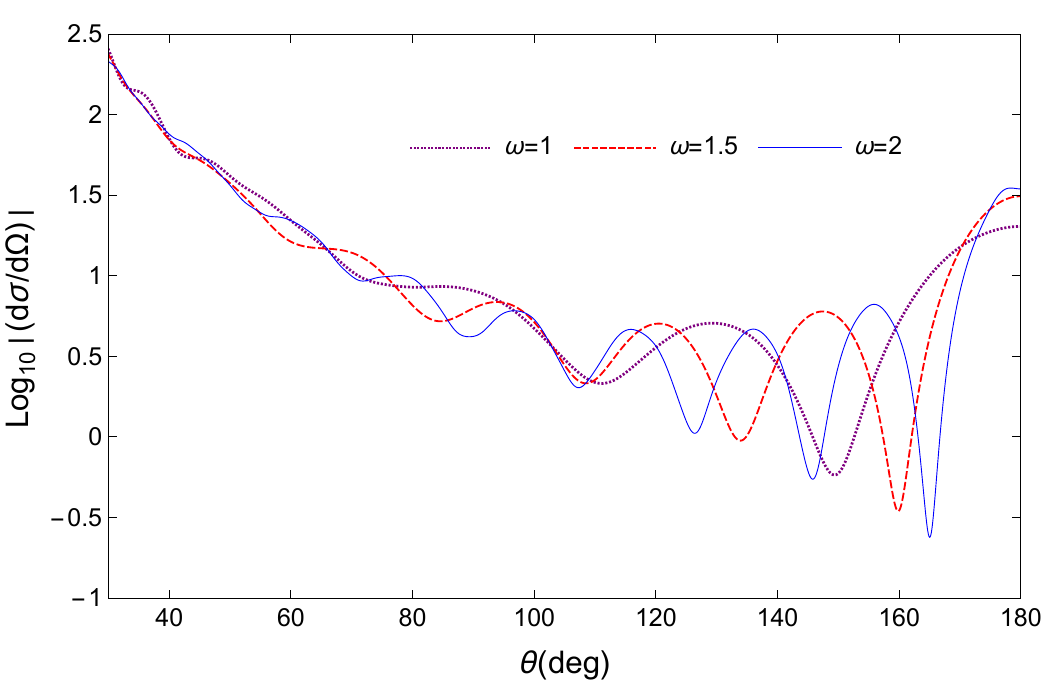}
\end{minipage}\hfill
\begin{minipage}{0.5\textwidth}
    \centering
    {\fontsize{8pt}{10pt}\selectfont $q=0.2$,$m=0.3$,$\omega=2$}\\
    \includegraphics[width=\linewidth, keepaspectratio]{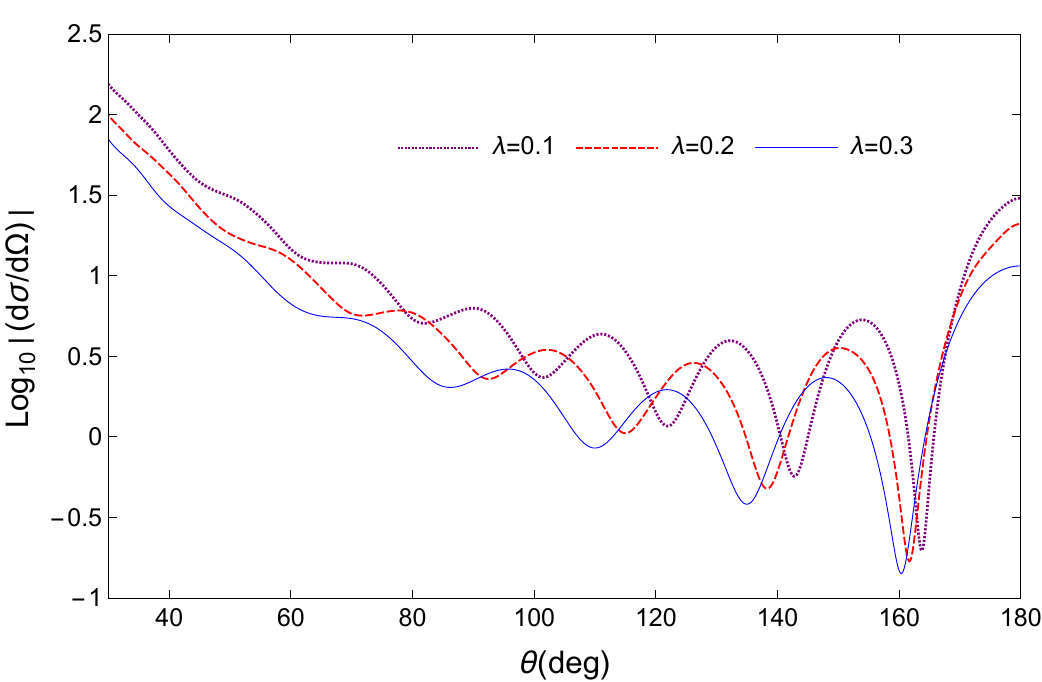}
\end{minipage}\hfill
\vspace{0.2cm}
\begin{minipage}{0.5\textwidth}
    \centering
    {\fontsize{8pt}{10pt}\selectfont $q=0.2$,$\omega=2$,$\lambda=0.2$}\\
    \includegraphics[width=\linewidth, keepaspectratio]{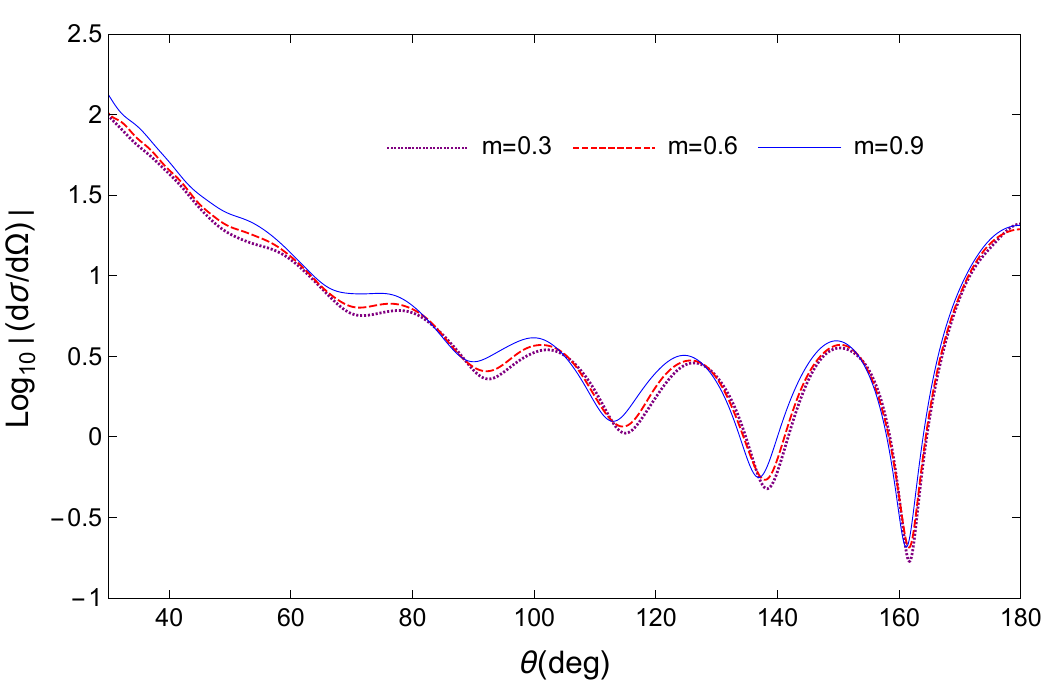}
\end{minipage}\hfill
\begin{minipage}{0.5\textwidth}
    \centering
    {\fontsize{8pt}{10pt}\selectfont $\omega=2$,$m=0.3$,$\lambda=0.3$}\\
    \includegraphics[width=\linewidth, keepaspectratio]{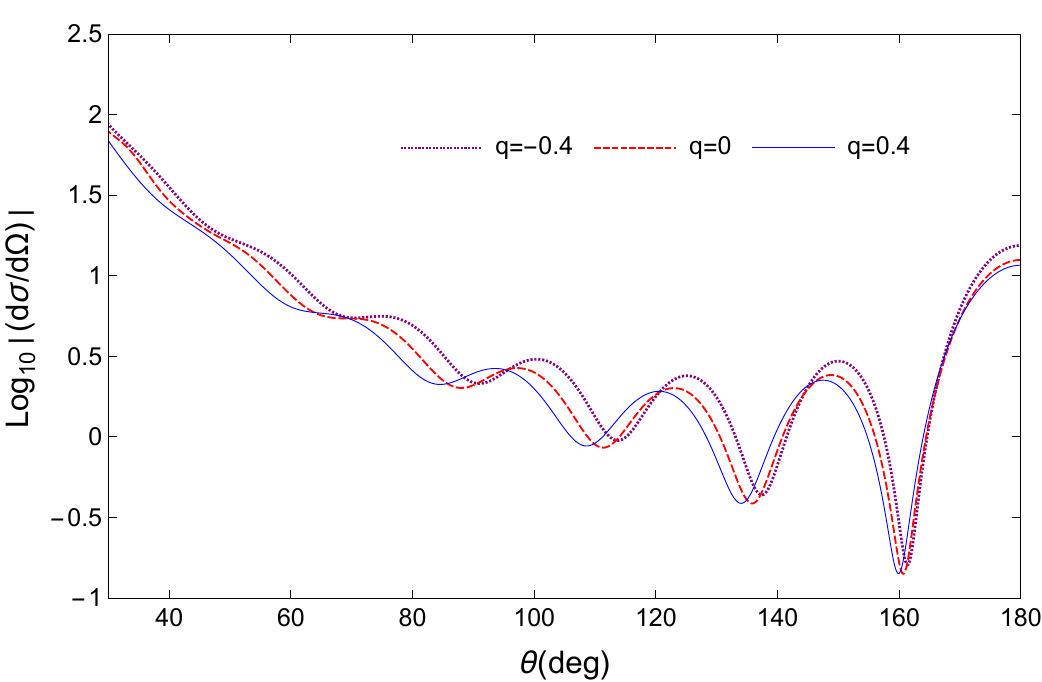}
\end{minipage}
\caption{\label{Differential} Scattering cross section for PFDM BHs with $M=1$, $Q=0.4$ for different values of $\omega$, $\lambda$, $m$ and $q$.}
\end{figure}

In Fig.~\ref{Differential}, we numerically study the differential scattering cross section of a charged massive particle propagating in the PFDM spacetime for various values of the parameters $\omega$, $\lambda$, $m$, and $q$. In the upper left panel, we present the variation of the differential scattering cross section with $\omega$. The parameters are set as $Q=0.4$, $q=0.2$, $m=0.3$, and $\lambda=0.05$. As can be seen, the differential scattering cross sections corresponding to small angles for different $\omega$ tend to be consistent, while they exhibit significant variations in the backward direction. As $\omega$ increases, the peaks of the interference fringes become higher and the troughs become lower accordingly. Moreover, the interference fringes in the large angle region are strongly dependent on the frequency $\omega$, lower frequencies correspond to wider fringe widths.

In the upper right panel, we present the effect of different $\lambda$ on the scattering cross section. The dark matter parameter $\lambda$ significantly modifies the scattering behavior. As $\lambda$ increases, the scattering cross section is suppressed overall. This behavior originates from the logarithmic term in the metric, which induces a long range modification of the effective potential. The logarithmic term in Eq.~(\ref{scattering second order}) is
\begin{equation}
-\frac{\lambda}{b}\left(1+\frac{1}{v^2}\right)\ln\frac{b}{2|\lambda|},
\end{equation}
which grows with $b$ due to the logarithmic factor. As a result, the overall deflection angle is reduced for larger $\lambda$, and the sensitivity of $\Theta(b)$ to $m$ becomes weaker. In the weak field limit, this leads to a correction in the deflection angle that reduces the overall bending, thereby suppressing large angle scattering. This behavior is fully consistent with the numerical results.

In the lower left panel, we present the variation of the differential scattering cross section with $m$. The parameters are set as $Q=0.4$, $q=0.2$, $\omega=2$, and $\lambda=0.2$. When the particle mass $m$ is increased, the scattering cross section increases slightly. When $\omega$ is fixed, a larger $m$ corresponds to a smaller particle velocity. A lower velocity leads to a stronger deflection of the particle in the long-range gravitational field and the dark matter modified field, and the differential scattering cross section consequently increases. Nevertheless, this increase is not drastic. For the adopted parameter $\omega=2$, the cases $m=0.3,0.6,0.9$ do not lie extremely close to the threshold.

In the lower right panel, we present the variation of the differential scattering cross section with $q$. The particle charge introduces an asymmetric effect on scattering. For $qQ>0$, leading to a suppression of the scattering cross section. In contrast, for $qQ<0$,  enhancing the scattering, particularly at large angles.

  In addition, we perform a comparison among three cases: classical scattering Eq.~(\ref{Classical  scattering}), glory scattering Eq.~(\ref{Glory scattering}), and numerical results of partial-wave scattering cross sections Eq.~(\ref{scattering cross section}) for different values of the parameter. The results are in excellent agreement, particularly in the backward direction.

\begin{figure}[!t]
\centering
\begin{minipage}{0.5\textwidth}
    \centering
    {\fontsize{8pt}{10pt}\selectfont $\omega=3$}\\
    \includegraphics[width=\linewidth, keepaspectratio]{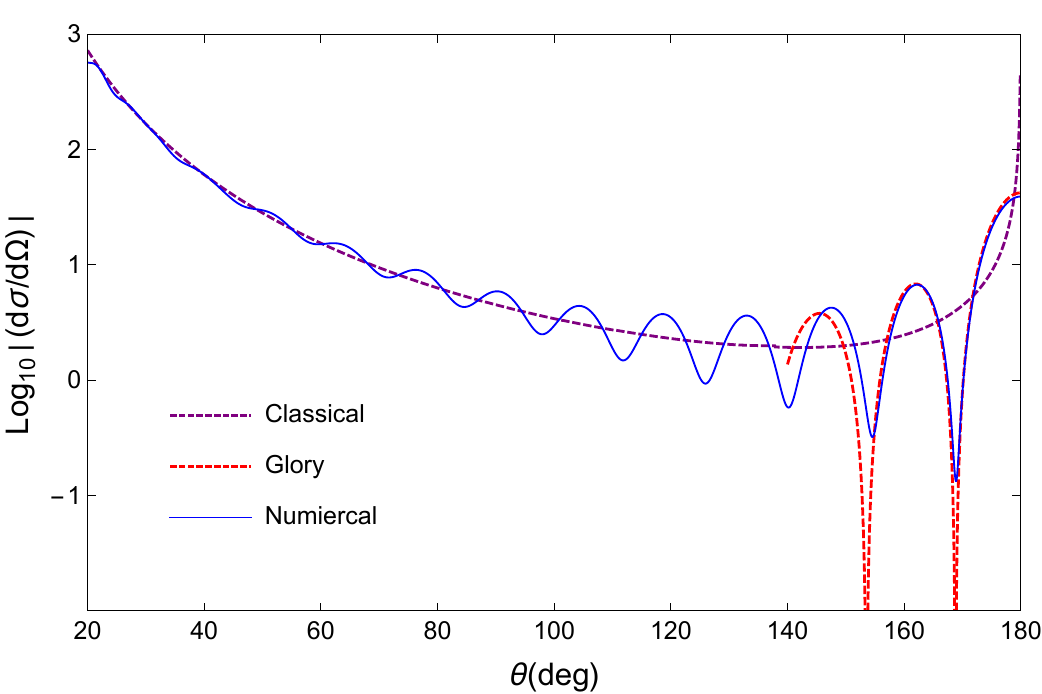}
\end{minipage}\hfill
\begin{minipage}{0.5\textwidth}
    \centering
    {\fontsize{8pt}{10pt}\selectfont $\omega=1$}\\
    \includegraphics[width=\linewidth, keepaspectratio]{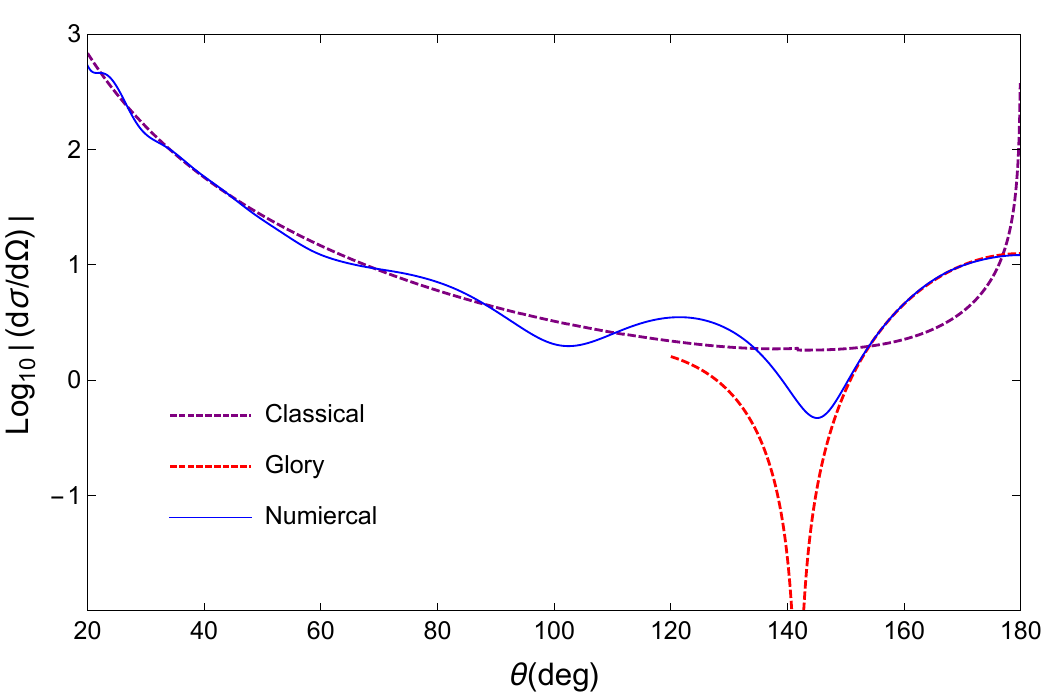}
\end{minipage}

\caption{\label{Differential2} Comparison of classical scattering, glory scattering and differential scattering cross section of PFDM BHs. We set $M=1$, $Q=0.5$, $q=0.3$, $m=0.3$, $\lambda=0.1$. The numerical results are in excellent agreement with those of classical scattering at small scattering angles. Meanwhile, in the backward scattering direction, the numerical results also match perfectly with glory scattering.}
\end{figure}
\section{Concluding remarks}

In this paper, we present an investigation of the scattering properties of a charged massive scalar field by an RN BH immersed in perfect fluid dark matter. We obtain high precision results for the absorption and scattering cross sections via the partial wave method. We discuss in detail the wave function equations in three regions and elaborate the relations among the various coefficients. To better understand the influence of dark matter on scattering, we derive the analytical expression for the scattering cross section in the weak field regime and compare it with numerical results.
From a physical perspective, dark matter affects the scattering process at multiple levels. In the asymptotic region, logarithmic terms modify the classical deflection angle as well as the relation between impact parameter and scattering angle. In the large angle regime, the interference between geometric optics trajectories and wave effects governs the structure of glory scattering, which is highly sensitive to the PFDM parameter $\lambda$. Therefore, scattering phenomena provide a direct and sensitive probe for dark matter effects in BH spacetimes.

It is worth mentioning the limiting behavior of the $\sigma_{abs}$ as $\omega$ approaches $m$. From Eq.~(\ref{geodesics}) and Eq.~(\ref{orbit1}), for massive particles in the limit $\omega\to m$, or equivalently $v\to0$, one obtains
\begin{equation}
\ell_c^2=\frac{r_c^2}{f(r_c)} \left[ \frac{A}{r_c} - \frac{B}{r_c^2} - \frac{\lambda}{r_c} \ln\left(\frac{r_c}{|\lambda|}\right) \right], \label{lc2}
\end{equation}
where 
\begin{equation}
\ell=L/m,\qquad
A = 2M - \frac{2qQ}{m},\qquad
B = Q^2 - \frac{q^2Q^2}{m^2}.
\end{equation}
When $\ell_c^2>0$, the capture condition is $\ell^2<\ell_c^2$, while $\ell^2>\ell_c^2$ corresponds to scattering.
It can be seen from Eq.~(\ref{lc2}) that the dominant term $A$ consists of the Newtonian term and the Coulomb force term. If $M>qQ/m$, the long range interaction is attractive. In the limit $\omega\to m$, the critical impact parameter grows without bound, so that the capture cross section diverges. Equivalently, particles with any fixed finite impact parameter are captured in this limit. If $M<qQ/m$, the Coulomb repulsion dominates the asymptotic interaction. In the limit $\omega\to m$, a particle released from infinity is repelled before reaching the BH, and the absorption cross section tends to zero. Only when $M\approx qQ/m$ do the terms $B$ and $\lambda$ become extremely important. In Fig.~\ref{absorption m lambda} and Fig.~\ref{absorption cross section q m lambda}, we can see when $q=-0.4$, $q=0$, the condition $M>qQ/m$ is satisfied, and the $\sigma_{abs}$ tends to diverge as $\omega\to m$. In Fig.~\ref{absorption cross section q m lambda}, when $q=1.2$, the condition $M<qQ/m$ is satisfied, for both the RN and PFDM BHs, $\sigma_{\rm abs}$ tends to zero in the low-frequency limit.

In terms of the differential scattering cross section, increasing $\omega$ leads to a significant reduction in the angular width of the fringes. This behavior can be understood from the glory scattering approximation Eq.~(\ref{Glory scattering}), where the oscillatory behavior is governed by the Bessel function. Near $\theta\approx\pi$, the angular spacing between fringes can be estimated as $\Delta\theta \sim \pi(\omega v b_g)^{-1}$. Therefore, higher frequencies correspond to shorter oscillation wavelengths in angular space, producing narrower and denser fringe patterns. Conversely, lower frequencies lead to broader and less pronounced fringes. In the region $\theta\ll1$, scattering is dominated by geometric scattering, whose leading term is $\textrm{d}\sigma/\textrm{d}\Omega\sim 1/\theta^4$. For $\lambda=0$ and $q=0$, the BH charge $Q$ leads only to a subdominant correction term proportional to $1/\theta^3$ at small angles\cite{Crispino2009}. For $\lambda=0$ but $q\neq0$ and $m\neq0$, from Eq.~(\ref{weak field expression2}), the correction to $q$, $Q$ and $m$ remains at order $1/\theta^4$ and appears in the form of a combination of $qQ/m$, which is slightly different from a charged particles in the charged Horndeski spacetime\cite{Li2025}. The parameters $\omega$, $q$, and $m$ only modify the preceding coefficient $K^2$ and do not alter the power-law exponent at small angles. 
From Eq.~(\ref{weak field expression}), when $\lambda\neq0$, the leading small angle behavior is logarithmically modulated and contains a term proportional to $\theta^{-4}\ln(1/\theta)$. Therefore, the small angle cross section no longer follows a pure power law. The $\theta^{-3}$ contribution arises only when the second-order deflection angle is fully retained. The parameter $qQ/m$ affects two components simultaneously: $K^2$ and $-\lambda K\ln|K|$. Consequently, the influence of $qQ/m$ is no longer limited to simply changing the amplitude of the cross section; it also modifies the logarithmic correction. Whether $\lambda$ vanishes or not, the qualitative role of $qQ/m$ remains the same. More specifically, when $qQ>0$, indicating that like charges repel each other. This weakens the effective gravitational focusing and reduces the small angle scattering cross section. In contrast, when $qQ<0$, indicating that opposite charges attract each other. This enhances the deflection and increases the small angle scattering cross section. However, for the parameter range considered here, where $K>0$, positive $\lambda$ gives a logarithmic suppression, as shown in Fig.~\ref{Differential}. For the parameter settings $q=0.2$, $m=0.3$, and $\omega=2$, an increase of $\lambda$ from $0.1$ to $0.3$ produces an overall suppression of the scattering cross section in the small angle region, whereas this effect is insignificant for other parameter combinations.

\acknowledgments

I would like to express my sincere gratitude to Professor Hong Lv and Professor Yi Pang of Tianjin University for their valuable comments and suggestions on this paper. This work is supported by the National Science Foundation of China under Grant Nos.12373022, U1731107, the Seventh Batch of High Level Innovative Talents of Guizhou Province under Grant No.GCC[2023]011, and the Special Funds for Discipline Construction and Postgraduate Education under Grant No.JX-2020-02.

\appendix
\section{Weak field deflection angle up to second order}

We consider the motion of a charged massive particle with mass \(m\) and charge \(q\) in the static spherically symmetric spacetime, define
\begin{equation}
\chi \equiv \frac{qQ\sqrt{1-v^2}}{m}, \qquad 
\zeta \equiv \frac{1-v^2}{v^2}.
\end{equation}
Then Eq.~(\ref{jiaodu}) becomes
\begin{equation}
\left(\frac{du}{d\varphi}\right)^2
=
\frac{(1-\chi u)^2}{b^2v^2}
-
\left(1-2Mu+Q^2u^2+\lambda u\ln\frac{1}{u|\lambda|}\right)
\left(\frac{\zeta}{b^2}+u^2\right).
\end{equation}
It is natural to separate the flat-space contribution from the gravitational, electromagnetic, and PFDM corrections. Expanding the above equation, we obtain
\begin{equation}
\left(\frac{du}{d\varphi}\right)^2
=
\frac{1}{b^2}-u^2+\delta\mathcal{T}(u),
\end{equation}
where the perturbation term is
\begin{equation}
\begin{aligned}
\delta\mathcal{T}(u)
={}&
-\frac{2\chi}{b^2v^2}u
+\frac{\chi^2}{b^2v^2}u^2
+2Mu\left(\frac{\zeta}{b^2}+u^2\right)
\\
&-Q^2u^2\left(\frac{\zeta}{b^2}+u^2\right)
-\lambda u\ln\frac{1}{u|\lambda|}
\left(\frac{\zeta}{b^2}+u^2\right).
\end{aligned}
\end{equation}
Introduce the small parameter $b_{\varepsilon}=1/b$ and rescale $u(\varphi)=b_{\varepsilon} y(\varphi)$. Then the equation reduces to
\begin{equation}
y''+y=b_{\varepsilon} F_1[y]+b_{\varepsilon}^2 F_2[y]+O(b_{\varepsilon}^3),
\end{equation}
where
\begin{equation}
\begin{aligned}
F_1[y]
={}&
-\frac{\chi}{v^2}
+M(\zeta+3y^2)
\\
&-\frac{\lambda}{2}\Big[
(\zeta+3y^2)\ln\frac{b}{|\lambda|y}
-(\zeta+y^2)
\Big],
\end{aligned}
\end{equation}
\begin{equation}
F_2[y]
=
\frac{\chi^2}{v^2}y
-\zeta Q^2 y
-2Q^2y^3.
\end{equation}
We now solve the orbit equation perturbatively by expanding
\begin{equation}
y(\varphi)=y_0+b_{\varepsilon} y_1+b_{\varepsilon}^2 y_2+\cdots.
\end{equation}
At zeroth order, the equation is simply
\begin{equation}
y_0''+y_0=0, \qquad y_0=\sin\varphi.
\end{equation}
At first order, one obtains
\begin{equation}
y_1''+y_1=F_1[y_0].
\end{equation}
The solution can be decomposed as
\begin{equation}
y_1=M y_M+\chi y_\chi+\lambda y_\lambda,
\end{equation}
with
\begin{equation}
y_M=\zeta(1-\cos\varphi)+(1-\cos\varphi)^2,
\end{equation}
\begin{equation}
y_\chi=-(\zeta+1)(1-\cos\varphi),
\end{equation}
\begin{equation}
\begin{aligned}
y_\lambda
={}&
-\frac{1}{2}y_M \ln\frac{b}{|\lambda|\sin\varphi}
\\
&+\frac{1}{2}(\zeta+2)\cos\varphi\ln\frac{1+\cos\varphi}{2}
-\frac{1}{2}\cos\varphi(\cos\varphi-1).
\end{aligned}
\end{equation}
At second order, the equation becomes
\begin{equation}
y_2''+y_2=
\left.\frac{\partial F_1}{\partial y}\right|_{y_0}y_1
+F_2[y_0],
\end{equation}
where
\begin{equation}
\left.\frac{\partial F_1}{\partial y}\right|_{y_0}
=
6M\sin\varphi
+\lambda\left[
\frac{\zeta}{2\sin\varphi}
+\frac{5}{2}\sin\varphi
-3\sin\varphi \ln\frac{b}{|\lambda|\sin\varphi}
\right].
\end{equation}
Using the perturbative orbit method, it admits the expansion
\begin{equation}
\Theta(b)=b_{\varepsilon}\Theta_1+b_{\varepsilon}^2\Theta_2+O(b_{\varepsilon}^3),
\end{equation}
with
\begin{equation}
\Theta_1=\int_0^\pi \sin\varphi\,F_1[y_0]\,d\varphi,
\end{equation}
\begin{equation}
\Theta_2=
\int_0^\pi \sin\varphi
\left[
\left.\frac{\partial F_1}{\partial y}\right|_{y_0}y_1
+F_2[y_0]
\right]d\varphi.
\end{equation}
The first-order result is
\begin{equation}
\Theta_1=
2M\left(1+\frac{1}{v^2}\right)
-\frac{2qQ\sqrt{1-v^2}}{mv^2}
-\lambda\left[
\left(1+\frac{1}{v^2}\right)\ln\frac{b}{2|\lambda|}
+1
\right].
\end{equation}
The second-order result can be written as
\begin{equation}
\Theta_2=
\Theta_{MM}+\Theta_{M\chi}+\Theta_{\chi\chi}
+\Theta_{Q^2}+\Theta_{M\lambda}
+\Theta_{\chi\lambda}+\Theta_{\lambda\lambda},
\end{equation}
where
\begin{equation}
\Theta_{MM}=\frac{3\pi}{4}(4\zeta+5)M^2,
\end{equation}
\begin{equation}
\Theta_{M\chi}=-3\pi(\zeta+1)M\chi,
\end{equation}
\begin{equation}
\Theta_{\chi\chi}=\frac{\pi\chi^2}{2v^2},
\end{equation}
\begin{equation}
\Theta_{Q^2}=-\frac{\pi}{4}(2\zeta+3)Q^2,
\end{equation}
\begin{equation}
\Theta_{M\lambda}
=
\pi M\lambda
\left[
\frac{4\zeta^2+32\zeta+31}{8}
-\frac{3}{4}(4\zeta+5)\ln\frac{2b}{|\lambda|}
\right],
\end{equation}
\begin{equation}
\Theta_{\chi\lambda}
=
-\pi(\zeta+1)\chi\lambda
\left[
\frac{\zeta}{2}+2-\frac{3}{2}\ln\frac{b}{2|\lambda|}
\right],
\end{equation}
\begin{equation}
\Theta_{\lambda\lambda}
=\pi\lambda^2\Bigg[
\frac{3}{16}(4\zeta+5)\ln^2\frac{2b}{|\lambda|}
-\frac{4\zeta^2+32\zeta+31}{16}\ln\frac{2b}{|\lambda|}
+\frac{\zeta^2}{4}+\frac{\zeta}{2}
+\frac{4\zeta+5}{64}\pi^2+\frac{5}{32}
\Bigg].
\end{equation}

\section{Classical differential scattering cross section}

Up to second order in the weak-field expansion, the deflection angle can be written as
\begin{equation}
\theta
=
\frac{1}{b}\Big[K-\lambda\beta\ln\frac{b}{2|\lambda|}-\lambda\Big]
+\frac{1}{b^2}\Big(B_0+B_1\ln b+B_2\ln^2 b\Big)
+O(b^{-3}),
\label{theta_expansion}
\end{equation}
where the coefficients are given by
\begin{equation}
\begin{aligned}
&B_{0}=\pi\Bigg[
\frac{3(4+v^{2})}{4v^{2}}M^{2}
-3M\varepsilon
+\frac{q^{2}Q^{2}(1-v^{2})}{2m^{2}v^{2}}
-\frac{2+v^{2}}{4v^{2}}Q^{2}
\\
&+M\lambda\left(\frac{3}{8}+\frac{3}{v^{2}}+\frac{1}{2v^{4}}-\frac{3(4+v^{2})}{4v^{2}}\ln\frac{2}{|\lambda|}\right)
-\varepsilon\lambda\left(\frac{1-v^{2}}{2v^{2}}+2+\frac{3}{2}\ln(2|\lambda|)\right)
\\
&+\lambda^{2}\Bigg(
\frac{3(4+v^{2})}{16v^{2}}\ln^{2}\frac{2}{|\lambda|}
-\frac{1}{16}\left(3+\frac{24}{v^{2}}+\frac{4}{v^{4}}\right)\ln\frac{2}{|\lambda|}
+\frac{1-v^{4}}{4v^{4}}
+\frac{4+v^{2}}{64v^{2}}\pi^{2}
+\frac{5}{32}
\Bigg)
\Bigg],
\end{aligned}
\end{equation}
\begin{equation}
B_{1}=\pi\Bigg[
-\frac{3(4+v^{2})}{4v^{2}}M\lambda
+\frac{3}{2}\varepsilon\lambda
+\lambda^{2}\left(
\frac{3(4+v^{2})}{8v^{2}}\ln\frac{2}{|\lambda|}
-\frac{1}{16}\left(3+\frac{24}{v^{2}}+\frac{4}{v^{4}}\right)
\right)
\Bigg],
\end{equation}

\begin{equation}
B_{2}=\pi\lambda^{2}\frac{3(4+v^{2})}{16v^{2}}.
\end{equation}

Due to the logarithmic term in Eq.~(\ref{theta_expansion}), the impact parameter cannot be expressed as a simple power of $\theta$. Instead, we introduce an effective function $\mathcal{A}(\theta)$ defined implicitly by
\begin{equation}
\mathcal{A}(\theta)
=
K-\lambda\beta\ln\frac{\mathcal{A}(\theta)}{2|\lambda|\theta}-\lambda.
\label{A_definition}
\end{equation}
At leading order, the inversion takes the form
\begin{equation}
b(\theta)
=
\frac{\mathcal{A}(\theta)}{\theta}
+
\frac{\mathcal{S}(\theta)}{\mathcal{A}(\theta)+\lambda\beta}
+O\!\left(\theta\,\ln^2\frac{1}{\theta}\right),
\label{b_theta}
\end{equation}
where we define
\begin{equation}
U(\theta)\equiv \ln\frac{\mathcal{A}(\theta)}{\theta},
\qquad
\mathcal{S}(\theta)=B_0+B_1 U(\theta)+B_2 U(\theta)^2.
\end{equation}
In the small-angle limit $\theta\ll 1$, using $\sin\theta\simeq\theta$, we obtain the asymptotic expansion
\begin{equation}
\begin{aligned}
\frac{d\sigma}{d\Omega}
=&\frac{\mathcal{A}(\theta)^3}{\big(\mathcal{A}(\theta)+\lambda\beta\big)\,\theta^4}
+\frac{\mathcal{A}(\theta)^2}{\big(\mathcal{A}(\theta)+\lambda\beta\big)^3\,\theta^3}
\\
&\Big[
\big(\mathcal{A}(\theta)+2\lambda\beta\big)\mathcal{S}(\theta)
+\big(\mathcal{A}(\theta)+\lambda\beta\big)\big(B_1+2B_2U(\theta)\big)
\Big].
\end{aligned}
\label{cross_section_final}
\end{equation}
Expanding Eq.~(\ref{A_definition}) to first order in $\lambda$, one finds
\begin{equation}
\mathcal{A}(\theta)
=K-\lambda\beta\ln\frac{|K|}{2|\lambda|\theta}-\lambda+O(\lambda^2).
\end{equation}
Substituting into Eq.~(\ref{cross_section_final}) , the leading contribution becomes
\begin{equation}
\frac{d\sigma}{d\Omega}
=
\frac{1}{\theta^4}
\left[
K^2
-\lambda K\left(
2\beta\ln\frac{|K|}{2|\lambda|\theta}
+\beta+2
\right)
\right]
+O\!\left(\theta^{-3},\lambda^2\right).
\end{equation}
This result shows that the logarithmic correction appears at order $\theta^{-4}$.

\end{document}